\documentclass[pdflatex,sn-mathphys-num]{sn-jnl}

\usepackage{graphicx}
\usepackage{amsmath,amssymb,amsfonts}
\usepackage{booktabs,array,tabularx}
\usepackage{xcolor}
\usepackage{placeins}
\usepackage{float}
\usepackage{fancyhdr}
\usepackage[UTF8,scheme=plain,fontset=none]{ctex}
\usepackage[most]{tcolorbox}

\newcolumntype{Y}{>{\raggedright\arraybackslash}X}
\newcommand{\wn}{\ensuremath{\mathrm{cm}^{-1}}}
\newcommand{\angstrom}{\text{\AA}}

\newcommand{\mainfigref}[1]{\ref{#1}}

\newtcolorbox{highlightbox}{
    breakable,
    colback=yellow!25,
    colframe=yellow!25,
    boxrule=0pt,
    arc=0pt,
    left=2pt,
    right=2pt,
    top=2pt,
    bottom=2pt,
    before skip=4pt,
    after skip=4pt
}

\graphicspath{{figures/}}
\renewcommand{\refname}{References}
\renewcommand{\abstractname}{Abstract}
\renewcommand{\headrulewidth}{0.4pt}

\begin{document}
\title[HARMONY]{HARMONY: an efficient and accurate deep learning framework for prediction of experimental vibrational spectra}
\author[1, 2]{\fnm{Xinyu} \sur{Lu}}\email{xinyulu@stu.xmu.edu.cn}
\author[2, 3]{\fnm{Bowen} \sur{Li}}
\author[1]{\fnm{Yi} \sur{Rong}} 
\author[1]{\fnm{Qingling} \sur{Zhou}} 
\author[1]{\fnm{Xingbo} \sur{Tang}} 
\author[1]{\fnm{Menghan} \sur{Li}} 
\author[4]{\fnm{Yuqiang} \sur{Li}}
\author[5]{\fnm{Guokun} \sur{Liu}}
\author*[2, 3]{\fnm{Tong} \sur{Zhu}}\email{tzhu@lps.ecnu.edu.cn}
\author*[1, 6]{\fnm{Bin} \sur{Ren}}\email{bren@xmu.edu.cn}
\author*[1, 6]{\fnm{Hao} \sur{Ma}}\email{oaham@xmu.edu.cn}

\affil[1]{\orgdiv{College of Chemistry and Chemical Engineering}, \orgname{Xiamen University}, \orgaddress{\city{Xiamen}, \postcode{361005}, \state{Fujian}, \country{China}}}

\affil[2]{\orgname{Shanghai Innovation Institute}, \orgaddress{\city{Shanghai}, \postcode{200030}, \country{China}}}

\affil[3]{\orgdiv{Shanghai Frontiers Science Center of Molecule Intelligent Syntheses, SII-ECNU Joint Autonomous Chemistry Laboratory, School of Chemistry and Molecular Engineering}, \orgname{East China Normal University}, \orgaddress{\city{Shanghai}, \postcode{200062}, \country{China}}}

\affil[4]{\orgname{Shanghai Artificial Intelligence Laboratory}, \orgaddress{\city{Shanghai}, \postcode{200000}, \country{China}}}

\affil[5]{\orgdiv{College of the Environment and Ecology}, \orgname{Xiamen University}, \orgaddress{\city{Xiamen}, \postcode{361005}, \state{Fujian}, \country{China}}}

\affil[6]{\orgdiv{Institute of Artificial Intelligence}, \orgname{Xiamen University}, \orgaddress{\city{Xiamen}, \postcode{361005}, \state{Fujian}, \country{China}}}

\abstract{
Predicting experimental vibrational spectra requires both efficient molecular calculations and a reliable mapping from theoretical vibrations to measured signals. Here we introduce \textbf{HARMONY}, a two-stage deep-learning framework that bridges this gap using harmonic spectra as physical anchors. Equivariant neural networks directly predict atomic forces and spectroscopic tensors for rapid harmonic-spectrum construction, while HarmoNet combines these spectra with molecular topology and measurement metadata to predict experimental infrared and Raman spectra. HARMONY achieves mean cosine similarities of 0.912 for infrared and 0.907 for Raman spectra, with a mean inference time of 2.12~s per molecule, approximately 600-fold faster than the evaluated DFT workflow. Beyond forward prediction, HARMONY improves inverse molecular inference through pseudo-experimental pretraining and spectral verification of candidate structures. Integrating HARMONY and Vib2Mol as scientific tools within a general-purpose agent enables adaptive candidate generation and verification. With molecular formulas provided, the resulting Vib2Mol--HARMONY Agent achieves Top-1 recall of 70.53\% on NIST using infrared spectra and 83.33\% on SDBS using combined infrared and Raman spectra. Fine-tuning with 19 amino-acid controls and eight peptides further increases mean Raman cosine similarity from 0.669 to 0.896 on 15 held-out peptides. HARMONY establishes harmonic anchoring as a practical connection between molecular physics, experimental spectroscopy and adaptive molecular structure elucidation.
}

\maketitle
\section{Introduction}\label{sec:introduction}

Vibrational spectroscopy, including infrared (IR) and Raman spectroscopy, provides characteristic molecular fingerprints that reveal chemical structure, bonding, and intermolecular interactions across chemical, materials, and biological systems \cite{wilson1980,morais2020}. Rapid and accurate prediction of experimental vibrational spectra directly from molecular structures would enable large-scale spectral assignment, molecular identification, and structural characterization.


Computational methods for predicting vibrational spectra range from quantum-chemical calculations to data-driven models. Quantum-chemical calculations derive vibrational spectra from normal modes and molecular response derivatives \cite{wilson1980}. Equivariant neural networks can now predict these quantities at substantially reduced computational cost \cite{detanet,xpainn}. Molecular dynamics captures finite-temperature molecular motion beyond the harmonic approximation. Combined with machine-learned interatomic potentials and response models, it enables spectral simulations over longer time scales and for larger systems \cite{gastegger2017,detanetmd,mace4ir}. Other data-driven models predict experimental spectra directly from molecular structure, sometimes incorporating measurement conditions, without explicit vibrational simulations \cite{chemprop,specgnn,pasgen}. Despite these advances, accurately and efficiently predicting experimental vibrational spectra remains challenging across diverse molecules and measurement conditions.

This challenge extends beyond computational efficiency to a persistent theory-to-experiment gap. This gap arises from limitations in molecular modeling and ensemble sampling, together with environmental and instrumental effects on measured spectra \cite{merrick2007,alberts2024}. Accelerating quantum-chemical calculations alone therefore does not ensure agreement with experiment, while direct structure-based prediction generally bypasses explicit vibrational representations. Learning a theory-to-experiment spectral mapping has also been explored for IR-based structure elucidation\cite{tao2026spec2spec}. This motivates retaining the physical information provided by quantum-chemical calculations and using neural networks to learn its relationship to experimental spectra. Specifically, harmonic spectra offer a natural anchor for this integration. Derived from the local curvature of a quantum-chemical potential-energy surface and molecular response derivatives, they provide a compact, frequency-resolved representation of molecular vibrations \cite{wilson1980}. Even when harmonic spectra differ substantially from experiment, they may provide complementary physical information beyond molecular topology and measurement metadata. Conditioned on molecular topology and measurement metadata, a learned mapping can translate harmonic representations into experimental spectra without explicitly simulating the full complexity of experimental conditions.


Here we introduce HARMONY (Harmonic-Anchored fRamework for Molecular vibratiONal spectroscopY prediction), a two-stage framework that employs harmonic spectra as physical anchors for predicting experimental IR and Raman spectra. HARMONY first employs equivariant neural networks to construct harmonic spectra from optimized molecular geometries, Hessians and spectroscopic response derivatives. It then learns to transform harmonic spectra into experimental spectra, conditioned on molecular topology and measurement metadata, thereby bridging the theory-experiment gap. Across experimental benchmarks, HARMONY achieves mean cosine similarities of 0.912 for infrared and 0.907 for Raman spectra, with second-scale inference and approximately 600-fold lower latency than the evaluated DFT workflow. These results support the use of harmonic spectra as physical anchors for learning experimental spectral features without explicitly simulating the full complexity of molecular dynamics and measurement effects. Beyond forward prediction, HARMONY enhances molecular structure elucidation by providing pseudo-experimental spectra for Vib2Mol pretraining and enabling spectral verification of candidate structures. Taking this integration further, we develop the Vib2Mol-HARMONY Agent by integrating both models as scientific tools within Codex, enabling adaptive candidate generation and spectral verification. With molecular formulas provided, the agent achieves Top-1 recall of 70.53\% on NIST using IR spectra and 83.33\% on SDBS using combined IR--Raman spectra. These results demonstrate how accurate forward spectral prediction can strengthen inverse molecular inference and, when integrated with general-purpose agents, enable more flexible molecular structure elucidation. Furthermore, adaptation with limited experimental data extends HARMONY to Raman prediction for short peptides.


\section{Results}\label{sec:results}
\subsection{A harmonic-anchored framework for experimental spectrum prediction}

\begin{figure}[htbp]
\centering
\includegraphics[width=\linewidth]{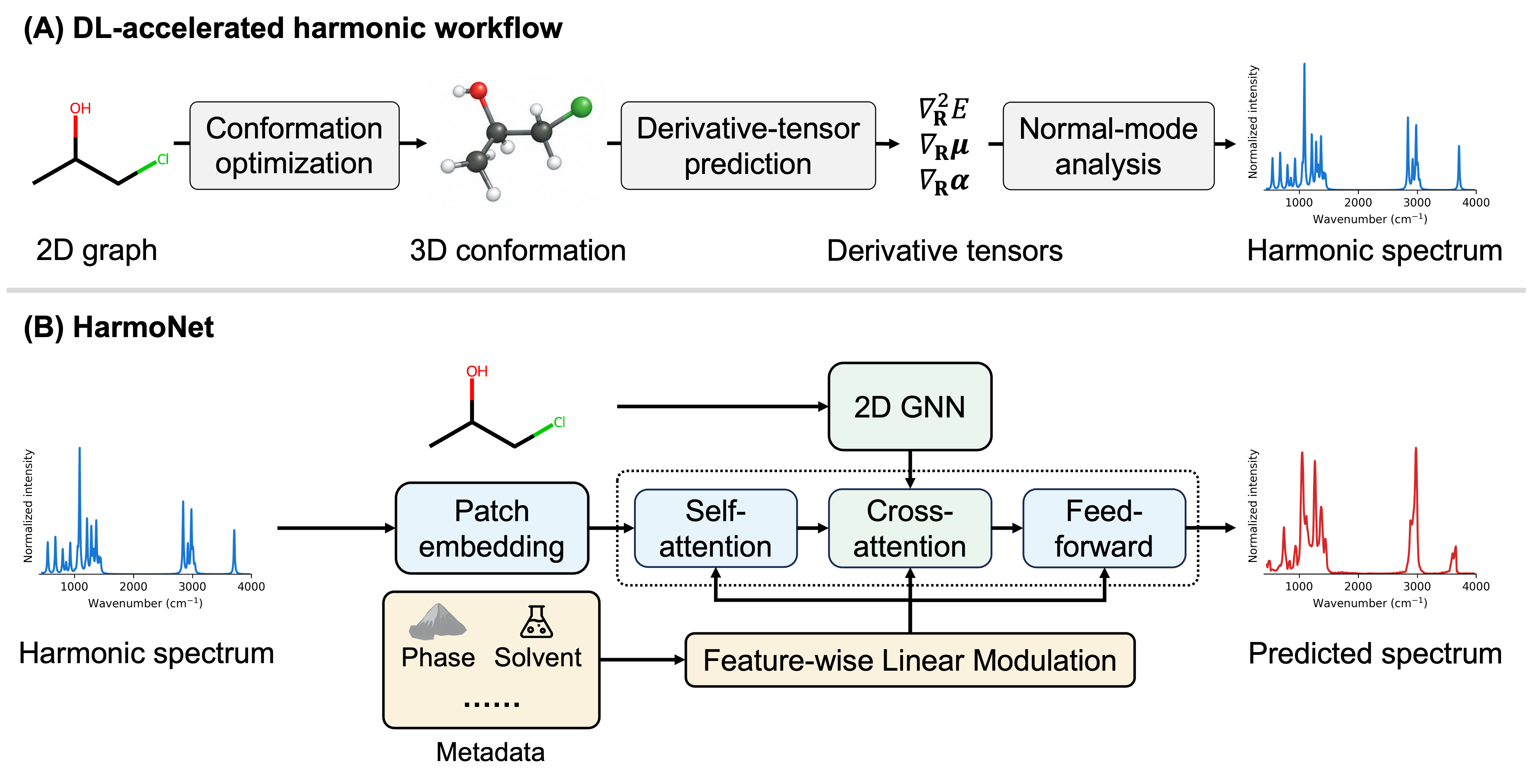}
\caption{\textbf{Two-stage architecture of HARMONY.} The first stage generates and optimizes a 3D molecular conformation from the 2D graph, then directly predicts the Hessian, dipole-derivative and polarizability-derivative tensors at the optimized geometry. Normal-mode analysis subsequently converts these quantities into harmonic IR and Raman spectra. The second stage processes harmonic-spectrum patches through six Transformer blocks combining spectral self-attention, molecular cross-attention and feed-forward layers. Adaptive layer normalization incorporates spectral modality and measurement metadata into HarmoNet, enabling the prediction of experimental IR and Raman spectra.}\label{fig:workflow}\label{fig:architecture}
\end{figure}

To bridge the theory–experiment gap, HARMONY employs a two-stage framework comprising deep-learning-accelerated harmonic-spectrum generation and a data-driven theory-to-experiment mapping implemented by HarmoNet (Fig.~\ref{fig:workflow}). In the first stage, a molecular conformer is generated and refined using a molecular mechanics force field, followed by geometry optimization using predicted atomic forces. Equivariant neural networks directly predict the atomic forces, Hessian, dipole derivatives and polarizability derivatives. Normal-mode analysis then converts these quantities into harmonic IR and Raman spectra. By avoiding automatic differentiation with respect to atomic coordinates, this workflow reduces computational cost while maintaining competitive accuracy in the predicted physical quantities. In the second stage, HarmoNet learns to map harmonic spectra to experimental spectra using molecular topology and available measurement metadata.

Within HarmoNet (Fig.~\ref{fig:workflow}B), each harmonic spectrum is partitioned into non-overlapping patches and embedded into spectral tokens. Six customized Transformer-like blocks then progressively transform the harmonic representation toward the experimental spectrum. In each block, measurement metadata, including sample phase, preparation method and matrix information, modulate the spectral-token representations through adaptive layer normalization. In parallel, molecular topology is encoded by a two-dimensional graph neural network (2D GNN) and incorporated into the spectral tokens through molecular cross-attention. Modality conditioning enables HarmoNet to learn from IR and Raman spectra jointly during training and produce modality-specific predictions at inference. The harmonic spectrum provides frequency-resolved vibrational information, while the molecular graph supplies chemical and topological context that cannot always be inferred from the spectrum alone. Measurement metadata further describe the sample environment and measurement conditions that shape the observed signal. The complementary contributions of these three information sources are systematically evaluated in a subsequent section.

HARMONY is trained in two stages using complementary quantum-chemical and experimental datasets. In the first stage, EquiformerV3-based models for geometry relaxation and spectroscopic-tensor prediction are trained on QMe14S, using over 500,000 single-point configurations and approximately 170,000 optimized geometries, respectively\cite{equiformerv3,qme14s}. In the second stage, HarmoNet learns the theory-to-experiment mapping using HarmoSpec, a dataset developed in this work that contains experimental IR and Raman spectra, molecular identities and associated metadata curated from NIST and SDBS\cite{nist,sdbs}. These experimental records are paired with harmonic spectra generated by the pretrained first-stage models. Dataset composition and split statistics are summarized in Supplementary Tables~\ref{tab:dataset} and \ref{tab:quantumdata}; curation and split construction are detailed in Supplementary Section~\ref{app:experimental-curation}.

\FloatBarrier
\subsection{Accurate prediction of experimental vibrational spectra with second-scale inference}

\begin{figure}[!htbp]
\centering
\includegraphics[width=\linewidth,height=\textheight,keepaspectratio]{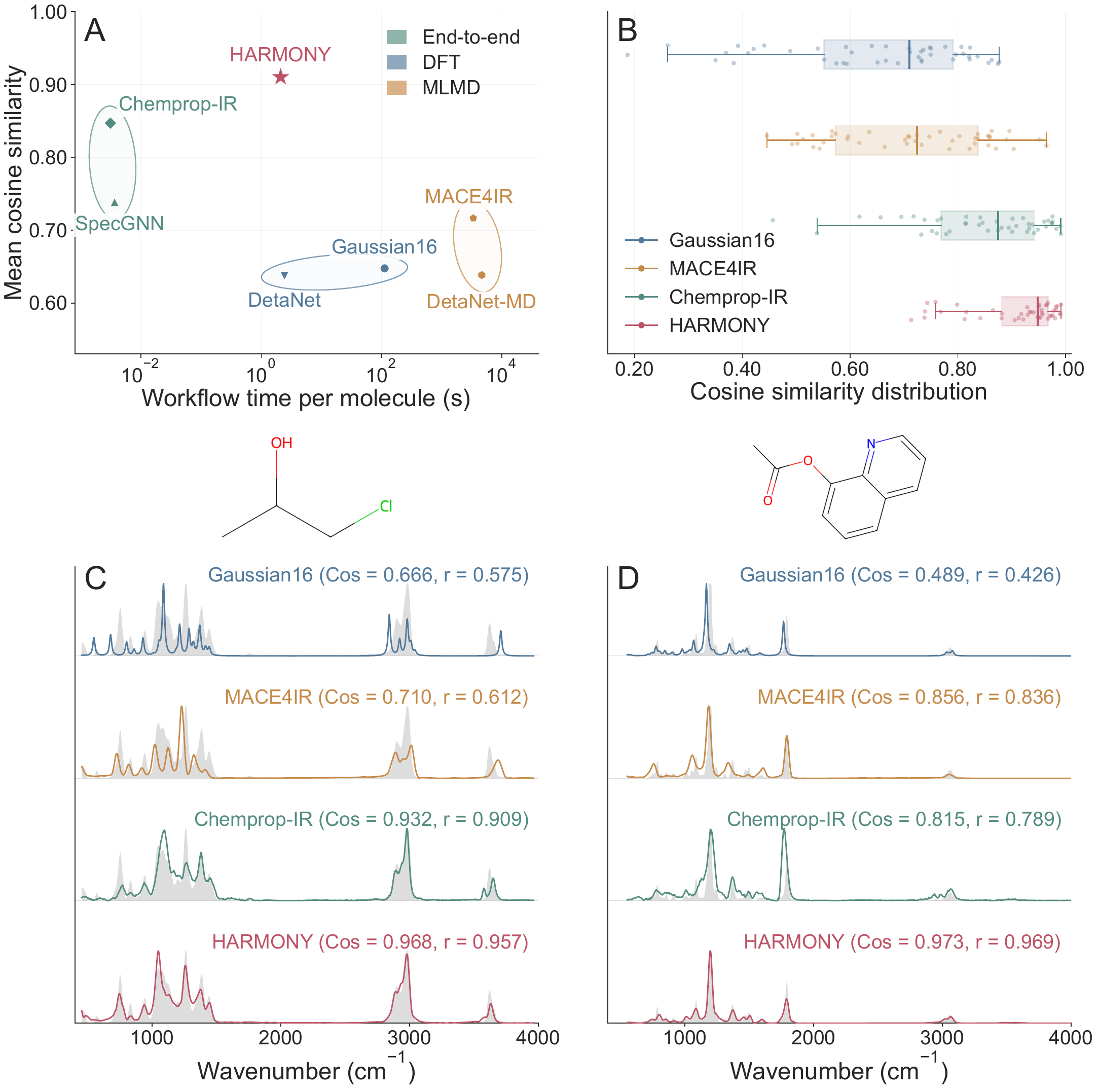}
\caption{(A) Mean cosine similarity versus per-molecule inference latency. Gaussian 16 uses CPU timings, while learned-model timings exclude loading and startup overhead. (B) Distributions of cosine similarity across the same 45 molecules. Points represent individual molecules, boxes indicate interquartile ranges, center lines denote medians, and whiskers extend to observations within 1.5 times the interquartile range. (C,D) Experimental IR spectra (gray shading) and corresponding predictions (colored lines) for (C) 1-chloropropan-2-ol and (D) quinolin-8-yl acetate.}\label{fig:accuracy}
\end{figure}

To compare predictive performance and computational cost across spectroscopic workflows, we selected a fixed panel of 45 molecules with experimental gas-phase IR spectra from the HarmoSpec test set, spanning a range of molecular sizes. We evaluated HARMONY against end-to-end neural predictors (SpecGNN\cite{specgnn} and Chemprop-IR\cite{chemprop}), harmonic-spectrum workflows based on quantum chemistry (Gaussian 16\cite{gaussian16}) or a learned DFT surrogate (DetaNet\cite{detanet}), and machine-learning molecular dynamics (MLMD; MACE4IR\cite{mace4ir} and DetaNet-MD\cite{detanetmd}). As shown in Fig.~\ref{fig:accuracy}A, HARMONY achieves the highest mean cosine similarity among the evaluated workflows, reaching 0.910 compared with 0.847 for Chemprop-IR, 0.648 for Gaussian 16 and 0.717 for MACE4IR. In terms of computational efficiency, end-to-end predictors remain the fastest, requiring only 3.1--3.7~ms per molecule. HARMONY and the DetaNet harmonic workflow require 2.10 and 2.43~s, respectively, compared with 112~s for Gaussian 16 and approximately 3,340--4,620~s for the tested MLMD workflows. HARMONY achieves the highest spectral similarity among the evaluated methods while maintaining second-scale inference, demonstrating a favorable performance--efficiency trade-off. Figure~\ref{fig:accuracy}B compares the distributions of cosine similarity between predicted and experimental spectra across the evaluated workflows. HARMONY shows consistently high similarity across the 45-molecule panel, rather than improvements driven by a few exceptionally well-predicted cases. Its sample standard deviation is 0.082, lower than those of Chemprop-IR (0.124), MACE4IR (0.145) and Gaussian 16 (0.177), indicating less variation in predictive accuracy across molecules. Molecule-selection criteria, size-stratified sampling and workflow timing protocols are detailed in Supplementary Section~\ref{app:workflow-timing}.

To illustrate these improvements at the spectral level, Fig.~\ref{fig:accuracy}C,D presents two selected examples, 1-chloropropan-2-ol and quinolin-8-yl acetate. HARMONY closely reproduces their experimental IR spectra, achieving cosine similarities of 0.968 and 0.973, respectively. In both cases, it captures the major band positions and relative intensities across 800--1800~\wn{} and, for 1-chloropropan-2-ol, the broad spectral envelope between 2800 and 3100~\wn. In contrast, harmonic DFT exhibits noticeable peak-position mismatch. For example, the dominant experimental peak near 1195~\wn{} in Fig.~\ref{fig:accuracy}D appears near 1167~\wn{} in the DFT spectrum. MACE4IR and Chemprop-IR also reproduce several prominent spectral features but show larger deviations in relative intensities and band profiles in these examples. These comparisons illustrate how the learned theory-to-experiment mapping improves agreement with experimental spectra beyond the harmonic approximation.

\FloatBarrier

\subsection{Complementary roles of harmonic spectra, molecular topology and measurement metadata}
\begin{figure}[!htbp]
\centering
\includegraphics[width=\linewidth]{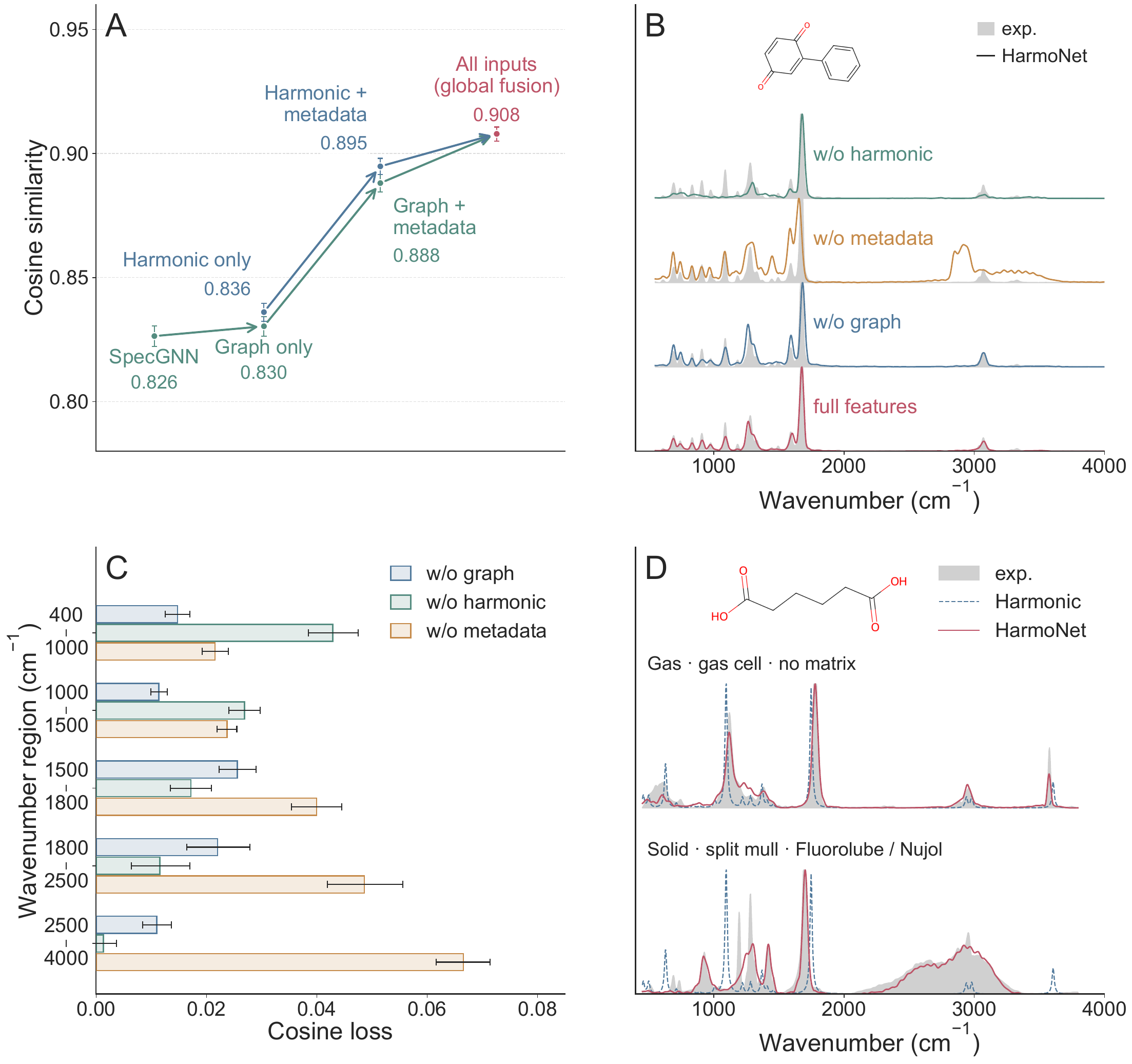}
\caption{\textbf{Complementary information in theory-to-experiment prediction.}
(A) IR prediction accuracy across combinations of harmonic information, measurement metadata and molecular topology, with SG(14,3) postprocessing. The three-input endpoint uses global graph conditioning; the final all-heavy-node cross-attention model is compared in Supplementary Table~\ref{tab:ablations}. Error bars indicate record-level bootstrap 95\% confidence intervals. (B) Input-ablation predictions for phenyl-substituted benzoquinone. (C) Regional loss in cosine similarity upon ablating harmonic, molecular topology or metadata information. (D) Metadata-conditioned predictions for adipic acid under gas-cell and solid split-mull conditions.}
\label{fig:ablation}
\end{figure}

HARMONY integrates harmonic spectra, molecular topology and measurement metadata. We assessed their contributions through input and fusion ablations on the HarmoSpec test set. For the SG(14,3)-smoothed predictions in Fig.~\ref{fig:ablation}A, graph-only HarmoNet achieves a mean IR cosine similarity of 0.830, compared with 0.826 for graph-only SpecGNN, while harmonic-only HarmoNet reaches 0.836. Adding measurement metadata increases these values to 0.888 and 0.895, respectively. Combining all three inputs through global graph conditioning yields 0.908. Replacing global conditioning with cross-attention to all heavy-atom nodes, the final HarmoNet architecture, raises cosine similarity to 0.911 using the same DFT-derived harmonic inputs and SG postprocessing (Supplementary Table~\ref{tab:ablations}). Together, these comparisons establish the complementary contributions of the three inputs and the additional benefit of node-level graph fusion. The corresponding Raman results are reported in Supplementary Table~\ref{tab:raman-ablation-trajectory}.

Figure~\ref{fig:ablation}B illustrates the contributions of the three information sources using phenyl-substituted benzoquinone as an example. Removing the harmonic input reduces agreement with several features in the 500--1000~\wn{} fingerprint region, while removing metadata introduces a broad feature above 2800~\wn{} that is absent from the experimental spectrum. Excluding molecular topology also alters the relative intensities of several bands. The model incorporating all features achieves the closest agreement with the experimental spectrum. To investigate these effects across the test set, we quantified the regional decrease in cosine similarity upon removing each information source (Fig.~\ref{fig:ablation}C). Harmonic information contributes most strongly in the 400--1000~\wn{} region, where its removal reduces cosine similarity by 0.043, while its contribution is relatively small above 2500~\wn. In contrast, metadata has the largest impact in the 2500--4000~\wn{} region, with a cosine decrease of 0.067, and also contributes appreciably between 1500 and 2500~\wn. Molecular topology provides benefits across all evaluated regions, with the largest observed gain of 0.026 between 1500 and 1800~\wn. Together, these results reveal distinct yet complementary roles for the three information sources. Harmonic spectra help resolve fine spectral features in the fingerprint region, measurement metadata contributes most strongly to high-wavenumber predictions, and molecular topology provides additional chemical context across the spectrum. To examine whether metadata remains useful without the high-wavenumber region, we further compared full-range training with training restricted to 400--2000~\wn. Both settings achieve comparable prediction accuracy within this region, while input ablations confirm that all three information sources remain beneficial (Supplementary Section~\ref{app:restricted-window} and Fig.~\ref{fig:restricted-window-ablation}). These findings indicate that the contribution of measurement metadata extends beyond high-wavenumber spectral features.

To test whether measurement metadata could distinguish conditions for the same molecule, we compared adipic acid measured in a gas cell and as a solid split mull (Fig.~\ref{fig:ablation}D). Starting from the same harmonic spectrum with a carbonyl-band maximum near 1748~\wn, HarmoNet predicts a blue shift to 1780~\wn{} for the gas-phase spectrum and a red shift to 1702~\wn{} for the solid-state spectrum. These predictions closely approach the corresponding experimental maxima at 1772 and 1692~\wn, respectively. Importantly, these condition-dependent spectral shifts are predicted without explicitly modeling intermolecular interactions, anharmonicity or environmental effects. Instead, HarmoNet learns the corresponding spectral corrections directly from experimental data, transforming the same harmonic representation into distinct experimental spectra through metadata-conditioned theory-to-experiment mapping.

We further assessed the contributions of individual metadata fields (Supplementary Tables~\ref{tab:ablations} and \ref{tab:raman-ablation-trajectory}; Fig.~\ref{fig:metadata-tree}). Phase information contributes most strongly to Raman prediction, whereas both phase and preparation substantially improve IR prediction, with matrix information providing only marginal additional gains.

\FloatBarrier

\subsection{Direct equivariant prediction accelerates harmonic-spectrum construction while preserving spectral fidelity}

Efficient construction of harmonic spectra is essential to the overall workflow of  HARMONY. To reduce computational cost, we therefore employ equivariant neural networks to directly predict atomic forces, Hessians, dipole derivatives and polarizability derivatives, rather than obtaining them through coordinate differentiation of learned energies or molecular response functions.

\begin{figure}[!htbp]
\centering
\includegraphics[width=\linewidth]{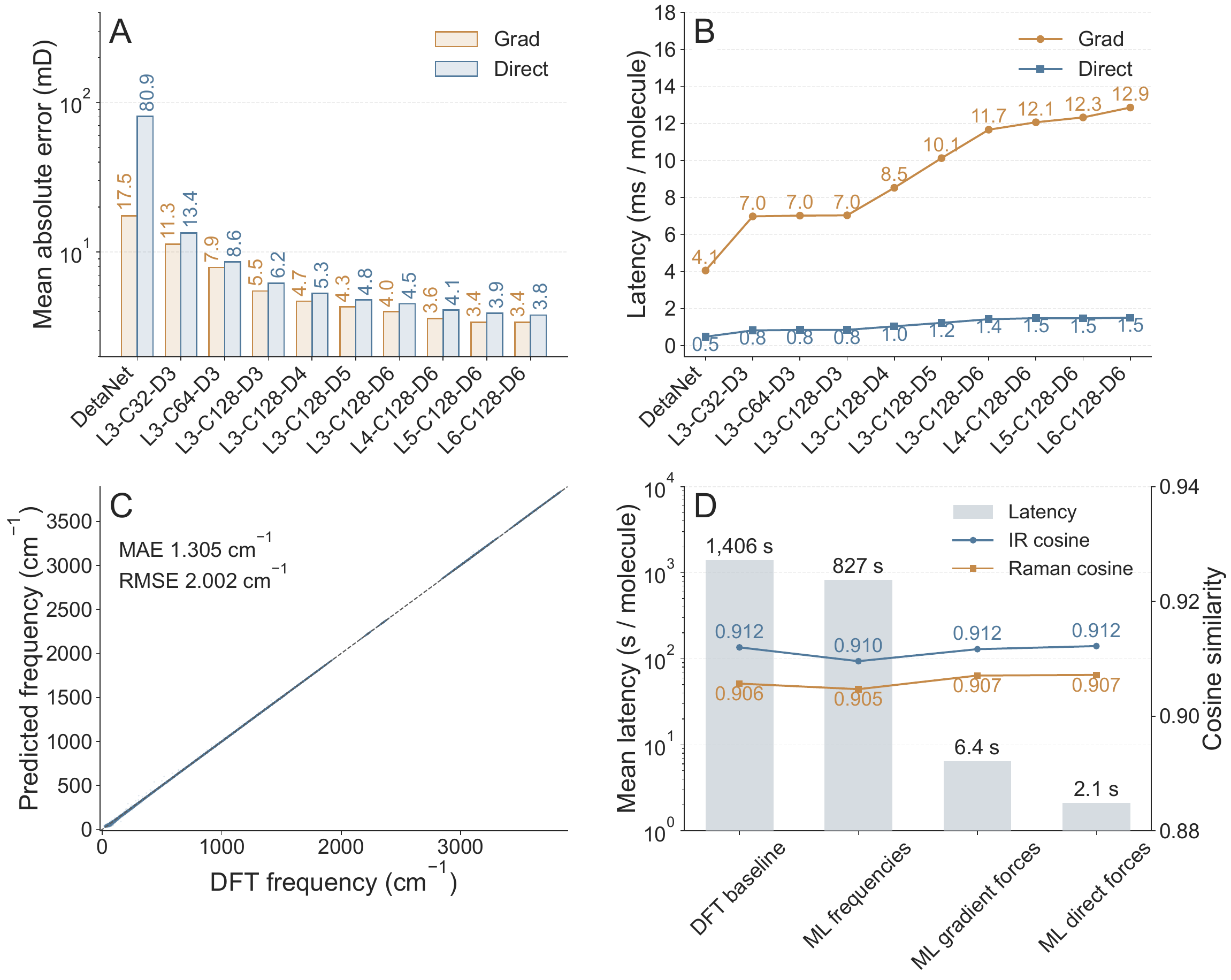}
\caption{(A) Dipole-derivative MAE for direct and differentiation-based prediction across DetaNet and EquiformerV3 configurations (log scale). (B) Per-molecule inference latency for the same configurations. (C) Predicted versus DFT vibrational frequencies for QM9S test molecules. (D) Progressive replacement of DFT stages. Bars show mean workflow latency across HarmoSpec test molecules (log scale), while lines indicate IR and Raman cosine similarities over experimental records, respectively.}
\label{fig:direct-harmonic}
\end{figure}

We first evaluated the accuracy--efficiency trade-off of direct dipole-derivative prediction on QM9S. We compared direct and differentiation-based approaches using DetaNet and EquiformerV3, varying the latter's maximum angular-momentum degree ($L$), channel width ($C$) and network depth ($D$). As shown in Fig.~\ref{fig:direct-harmonic}A, direct prediction yields substantially larger errors with DetaNet, whereas the accuracy gap generally narrows as EquiformerV3 becomes more expressive. For the largest EquiformerV3 configuration, direct prediction achieves comparable accuracy with approximately 12\% higher MAE, while reducing inference latency by 8.5-fold (Fig.~\ref{fig:direct-harmonic}B) and training cost by 9.8-fold (Supplementary Table~\ref{tab:fig4-training-costs}). Similar computational advantages extend to Hessians and polarizability derivatives, although the accuracy trade-offs vary across properties (Supplementary Table~\ref{tab:tensor-grad-direct}). Additional multi-property benchmarks on QM9S and QMe14S-SP are provided in Supplementary Tables~\ref{tab:qm9s-properties} and \ref{tab:qme14s-properties}. Together, these results suggest that greater backbone expressivity can reduce the reliance on differentiation-based inductive biases, allowing direct tensor prediction to achieve competitive accuracy at substantially lower computational cost.

We next examined whether direct tensor prediction compromises harmonic-spectrum fidelity. Rigid-body-projected analysis reveals imaginary internal modes in 3.87\% of QM9S test molecules, with 2.53\% and 0.65\% exhibiting frequencies below $-20$ and $-50$~\wn, respectively (Supplementary Table~\ref{tab:physical-diagnostics}). Nevertheless, as shown in Fig.~\ref{fig:direct-harmonic}C, directly predicted tensors yield low molecule-averaged vibrational-frequency errors (MAE = 1.305~\wn; RMSE = 2.002~\wn). The reconstructed harmonic IR and Raman spectra achieve mean cosine similarities of 0.987 and 0.977 against DFT references at the same molecular geometries (Supplementary Fig.~\ref{fig:harmonic-fidelity}). Notably, molecules exhibiting imaginary internal modes are retained in the evaluation and maintain high IR and Raman cosine similarities of 0.988 and 0.983, respectively. These results demonstrate that direct tensor prediction preserves high spectroscopic fidelity despite imperfect Hessian stability and the absence of strict derivative consistency. Additional frequency-error distributions and physical diagnostics are provided in Supplementary Tables~\ref{tab:spectral-fidelity}--\ref{tab:tensor-equivariance}.

Having established harmonic-spectrum fidelity at fixed geometries, we evaluated the complete HARMONY workflow by progressively replacing DFT geometry optimization and frequency calculations with DL-based models (Fig.~\ref{fig:direct-harmonic}D). The four configurations range from DFT-based harmonic-spectrum construction to a DL-driven workflow using directly predicted forces and spectroscopic tensors. Across the test set, mean latency decreases from approximately 1,406~s for the DFT workflow to 6.42~s using learned gradient-based forces, and further to 2.12~s per molecule with directly predicted forces. Direct-force geometry optimization converges for 99.92\% of molecules in HarmoSpec, with the remaining cases successfully optimized using energy-gradient relaxation (Supplementary Section~\ref{app:direct-force-validation} and Table~\ref{tab:direct-force-validation}). At this latency, the DL-driven workflow achieves mean cosine similarities of 0.912 for IR and 0.907 for Raman (Fig.~\ref{fig:direct-harmonic}D). Direct force and tensor prediction therefore reduce workflow latency while retaining experimental-spectrum accuracy.

To examine whether HarmoNet depends on the source of harmonic spectra, we evaluated the transferability of the learned mapping without retraining. HarmoNet trained on DL-predicted harmonic spectra achieves IR and Raman cosine similarities of 0.908 and 0.902 when supplied with DFT-derived inputs. Conversely, HarmoNet trained on DFT-derived spectra achieves corresponding similarities of 0.908 and 0.904 with DL-predicted inputs (Supplementary Fig.~\ref{fig:cross-route-progressive}). This bidirectional transfer suggests that HarmoNet learns a transferable relationship between local molecular vibrations and experimental spectra, rather than corrections specific to a particular harmonic-spectrum source. The harmonic spectrum thus serves as a physical anchor connecting the local curvature of the molecular potential-energy surface and spectroscopic response to experimental observations.

\subsection{Forward spectral prediction enhances molecular structure elucidation}

Structure-to-spectrum prediction and spectrum-to-structure inference represent complementary directions of the same spectrum--structure relationship. We investigate how accurate forward prediction can enhance inverse molecular inference through pseudo-experimental pretraining and spectral reranking, and further extend these capabilities through agent-assisted inference.

\begin{figure}[!htbp]
\centering
\includegraphics[width=.93\linewidth]{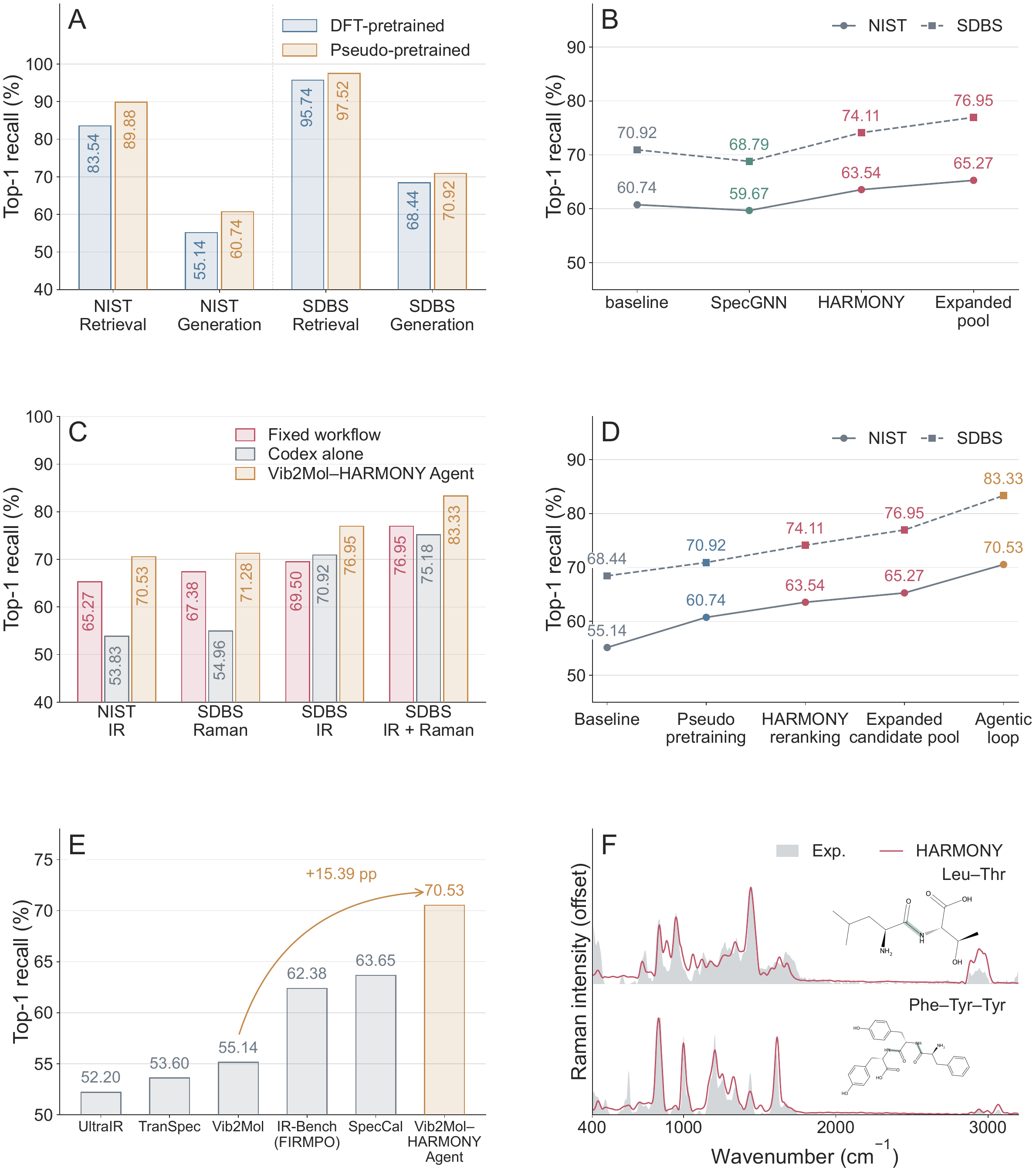}
\caption{\textbf{Applications of HARMONY in molecular structure elucidation and peptide Raman prediction.}
(A) Retrieval and generation performance on NIST and SDBS with DFT and pseudo-experimental pretraining.
(B) Top-1 recall with Vib2Mol matching, SpecGNN and HARMONY reranking, and expanded candidate pools.
(C) Top-1 recall of fixed workflows, Codex alone and the Vib2Mol--HARMONY Agent on NIST (IR) and SDBS (Raman, IR and combined IR--Raman).
(D) Performance progression through pseudo-experimental pretraining, HARMONY reranking, candidate-pool expansion and agent-assisted inference. All NIST stages use 1,215 test records; missing forward-reranking outputs count as incorrect. In (B,D), circles with solid lines represent NIST, and squares with dashed lines represent SDBS.
(E) Comparison of NIST IR Top-1 recall with UltraIR\cite{tan2026ultrair}, TranSpec\cite{specgnn}, Vib2Mol\cite{vib2mol}, IR-Bench (FIRMPO)\cite{tan2026firmpo} and SpecCal\cite{chen2026speccal}.
(F) Experimental Raman spectra (gray) and peptide-adapted HarmoNet predictions (red) for held-out Leu--Thr (LT) and Phe--Tyr--Tyr (FYY).}
\label{fig:inverse-enablement}
\end{figure}

The first strategy addresses the scarcity of experimental vibrational spectra. Theoretical spectra provide a valuable source of pretraining data for spectrum-to-structure models, but the theory--experiment gap can limit their effectiveness on experimental data \cite{vib2mol,alberts2024}. To address this limitation, we use HARMONY to generate pseudo-experimental IR and Raman spectra for Vib2Mol pretraining. We evaluated this approach on the original SDBS IR--Raman and NIST IR benchmarks, with molecular formulas provided as additional constraints. As shown in Fig.~\ref{fig:inverse-enablement}A, pseudo-experimental pretraining improves generation Top-1 recall from 68.44\% to 70.92\% on SDBS and from 55.14\% to 60.74\% on NIST. Retrieval Recall@1 also increases from 95.74\% to 97.52\% on SDBS and from 83.54\% to 89.88\% on NIST. Pseudo-experimental pretraining thus improves both molecular generation and retrieval on these benchmarks.

The second strategy uses HARMONY as a forward verification model to rerank candidate structures generated by Vib2Mol. For each candidate, HARMONY predicts the corresponding experimental spectrum, which is then compared with the measured spectrum to assess spectral consistency. On a fixed SDBS candidate pool, HARMONY reranking achieves a Top-1 recall of 74.11\%, compared with 68.79\% using SpecGNN (Fig.~\ref{fig:inverse-enablement}B). Similar improvements are observed on NIST, where Top-1 recall increases from 59.67\% with SpecGNN to 63.54\% with HARMONY. These results highlight the importance of accurate forward prediction for effective candidate reranking. However, reranking cannot recover structures absent from the candidate pool. We therefore combine the formula-filtered candidates from two Vib2Mol checkpoints and rerank them with HARMONY, achieving Top-1 recall of 76.95\% on SDBS and 65.27\% on NIST. This highlights the complementary benefits of candidate-pool expansion and accurate spectral verification.

Although candidate-pool expansion improves structure identification, fixed workflows remain limited by the structures already generated. To overcome this limitation, we integrate Vib2Mol and HARMONY as scientific tools into Codex, forming the Vib2Mol--HARMONY Agent for adaptive candidate generation and spectral verification. This design allows the agent to explore new structural hypotheses beyond a fixed candidate pool. We evaluated natural-language structure-elucidation questions from NIST (IR) and SDBS (Raman, IR or both), with molecular formulas provided in all cases (Fig.~\ref{fig:inverse-enablement}C). On NIST, the agent achieves a Top-1 recall of 70.53\%, compared with 53.83\% for Codex alone and 65.27\% for the fixed Vib2Mol--HARMONY workflow. On SDBS, the agent achieves 71.28\%, 76.95\% and 83.33\% for Raman-only, IR-only and combined IR--Raman inputs, respectively, outperforming both the corresponding fixed workflows (67.38\%, 69.50\% and 76.95\%) and Codex alone (54.96\%, 70.92\% and 75.18\%). With combined IR--Raman inputs, the agent identifies 30 correct structures absent from the fixed candidate pool, recovering 35 previously incorrect predictions while losing 17 correct ones. These results highlight how specialized spectroscopy tools complement general-purpose agents, extending molecular structure elucidation beyond fixed candidate pools. Paired outcomes and fixed-pool coverage for NIST and SDBS are reported in Supplementary Table~\ref{tab:agent-candidate-analysis}.

Figure~\ref{fig:inverse-enablement}D summarizes the progression from the original Vib2Mol baseline through pseudo-experimental pretraining, HARMONY reranking, candidate-pool expansion and agent-assisted inference. Across these configurations, Top-1 recall increases from 55.14\% to 70.53\% on NIST and from 68.44\% to 83.33\% on SDBS. Forward spectral prediction supports inverse molecular inference through improved training data, candidate verification and adaptive exploration. We further benchmarked the Vib2Mol--HARMONY Agent against published NIST IR structure-elucidation results (Fig.~\ref{fig:inverse-enablement}E). Its Top-1 recall of 70.53\% exceeds the reported values for UltraIR (52.20\%) \cite{tan2026ultrair}, TranSpec (53.60\%) \cite{specgnn}, Vib2Mol (55.14\%) \cite{vib2mol}, IR-Bench (FIRMPO; 62.38\%) \cite{tan2026firmpo} and Patch Transformer with SpecCal (63.65\%) \cite{chen2026speccal}. The agent exceeds the original Vib2Mol result by 15.39 percentage points.

\FloatBarrier

\subsection{Adaptation to peptide Raman spectra with limited experimental data}

Bottom-up protein analysis focuses on shorter peptide fragments, yet their sequence space remains vast: the 20 canonical amino acids permit $20^L$ sequences of length $L$, even without chemical modifications. Raman spectroscopy provides vibrational fingerprints for peptide characterization, but peptide-bond formation and conformational variability make their spectra more complex than combinations of isolated amino-acid signatures\cite{wei2008peptides}. Accurate sequence-to-spectrum prediction could therefore support peptide identification by providing reference spectra for candidate sequences. Here, we investigate whether HARMONY can be extended from small molecules to short peptides using limited experimental data.

To establish a DL-driven workflow for peptide Raman prediction, we first fine-tuned the equivariant models for geometry optimization and spectroscopic-tensor prediction on peptide quantum-chemical data. These models then generated harmonic Raman spectra from peptide structures without requiring DFT calculations at inference. The experimental dataset comprises 19 amino-acid controls and 25 peptide spectra measured from powder samples using 785-nm excitation. We used all 19 amino-acid controls and eight peptide spectra for training, two peptides for validation and 15 dipeptide and tripeptide sequences for testing. HarmoSpec-pretrained HarmoNet was subsequently fine-tuned on the 27 training spectra, with checkpoints selected based on validation cosine similarity (Supplementary Section~\ref{app:peptides}).

On the 15 held-out peptides, HarmoNet fine-tuning improves mean Raman cosine similarity from 0.669 to 0.896, compared with 0.497 for the harmonic inputs alone. All 15 test predictions improve after adaptation. Importantly, both pretrained and fine-tuned HarmoNet models use identical harmonic inputs, measurement metadata and experimental targets, isolating the improvement in the learned theory-to-experiment mapping. These results demonstrate that limited experimental peptide data can substantially improve spectral prediction while retaining a fully DL-driven harmonic-spectrum workflow.

Figure~\ref{fig:inverse-enablement}F illustrates the predicted Raman spectra of the held-out dipeptide Leu--Thr and tripeptide Phe--Tyr--Tyr, showing close agreement with their experimental profiles. Predictions for all 15 test peptides from the same checkpoint are provided in Supplementary Fig.~\ref{fig:peptides}. Together, these findings support the transferability of HARMONY from small molecules to short peptides and highlight its potential for future spectral verification of candidate peptide sequences.
\section{Discussion}\label{sec:discussion}

HARMONY connects molecular calculations with experimental vibrational spectroscopy through a two-stage framework anchored by harmonic spectra. Direct equivariant prediction enables efficient harmonic-spectrum construction, while HarmoNet combines these spectra with molecular topology and measurement metadata to learn the theory-to-experiment mapping. This approach achieves accurate experimental IR and Raman prediction with second-scale inference. Beyond forward prediction, pseudo-experimental pretraining and spectral verification improve molecular structure elucidation, while adaptation to short-peptide Raman spectra demonstrates the potential to extend HARMONY beyond its original small-molecule domain using limited experimental data.

A central insight is that the learned theory-to-experiment mapping does not require exact reproduction of DFT harmonic spectra. DFT and DL-driven workflows can produce different optimized geometries and local curvatures of the molecular potential-energy surface, yet HarmoNet maintains comparable predictive accuracy when harmonic spectra from either source are exchanged without retraining. This transfer suggests that the model relies on vibrational information shared across these computational approximations, rather than details specific to a particular harmonic-spectrum source. In this sense, HARMONY indirectly learns how the local curvature of the molecular potential-energy surface and spectroscopic response derivatives relate to experimental spectra through a shared harmonic representation.

The complementary contributions of harmonic spectra, molecular topology and measurement metadata further clarify how this mapping operates. Harmonic information contributes most strongly to fingerprint-region predictions, whereas metadata has its greatest impact at higher wavenumbers, potentially reflecting the sensitivity of these spectral features to sample environments and measurement conditions. Notably, metadata remains beneficial even when prediction is restricted to 400--2000~\wn. For adipic acid, HarmoNet predicts distinct carbonyl-band shifts under gas-cell and solid split-mull conditions from the same harmonic input, illustrating how experimental effects can be learned without explicitly simulating the surrounding environment. These findings highlight the value of combining physically grounded vibrational information with flexible, condition-dependent learning.

Beyond improving conventional structure-elucidation workflows, accurate forward spectral prediction also creates opportunities for agent-assisted molecular inference. Fixed candidate-generation and reranking workflows remain limited by the structures initially proposed, whereas general-purpose agents can adaptively explore and verify new structural hypotheses. By integrating Vib2Mol and HARMONY as scientific tools within Codex, the Vib2Mol--HARMONY Agent combines inverse structure generation with forward spectral verification. Its ability to identify correct structures beyond the fixed candidate pool demonstrates the complementary value of specialized scientific models and adaptive search. More broadly, this integration illustrates how forward models can evolve from standalone predictors into scientific tools that help general-purpose agents propose, evaluate and refine molecular hypotheses.

Nevertheless, agreement with experimental spectra does not necessarily establish the underlying microscopic mechanisms, including conformational populations, intermolecular interactions and anharmonic dynamics. High spectral similarity alone also cannot guarantee discrimination between closely related molecular structures. Future improvements could incorporate explicit conformational ensembles, broader coverage of chemical and measurement conditions, and calibrated prediction uncertainty to strengthen the reliability of spectral verification. For peptides, the present results motivate further investigation of whether forward spectral prediction can distinguish competing sequences with subtle structural differences. More broadly, combining rapid spectral prediction with adaptive molecular inference and experimental feedback offers a promising direction for molecular identification, spectral assignment and autonomous experimental planning.
\section{Methods}\label{sec:methods}

\subsection{Datasets and preprocessing}\label{sec:methods-data}

\paragraph{Quantum-chemical datasets.}

The production EquiformerV3 models were trained on QMe14S\cite{qme14s}. Single-point configurations provided supervision for energy and force prediction, while optimized geometries supplied Hessians, dipole derivatives and polarizability derivatives. Separate models were trained on QM9S\cite{detanet} to benchmark network architectures and evaluate the accuracy, computational efficiency and physical fidelity of direct prediction.

\paragraph{HarmoSpec construction.}

HarmoSpec combines experimental IR spectra from NIST and IR/Raman spectra from SDBS\cite{nist,sdbs}. Each record contains a molecular identity, an experimental spectrum and associated measurement metadata. Each experimental spectrum is paired with two harmonic representations, one obtained from DFT calculations and the other generated using pretrained equivariant models. For a given molecule and spectral modality, the harmonic input from each computational route remains fixed across different experimental conditions.

All experimental spectra were preprocessed on a common grid of 1,024 points spanning 400--4,000~\wn. Preprocessing included Savitzky--Golay smoothing (nine-point window, quadratic polynomial)\cite{savitzky1964}, adaptive Zhang baseline correction\cite{zhang2010airpls} and robust min--max normalization using the 0.5th and 99.5th intensity percentiles. Unmeasured spectral regions were explicitly masked rather than treated as zero intensity. Near-duplicate spectra were identified by spectral similarity within molecular and metadata groups, with higher signal-to-noise measurements preferentially retained. After curation and deduplication, HarmoSpec contains 33,423 IR and 2,522 Raman spectra.

\paragraph{Measurement metadata and molecular splits.}

Measurement metadata include sample phase, preparation method, matrix identities for high- and low-wavenumber regions, and the boundary between these regions. Metadata are associated with individual measurements, allowing the same molecule to have different experimental conditions.

To prevent molecular overlap across data splits, molecules were grouped by Bemis--Murcko scaffolds\cite{bemis1996}, retaining stereochemical information. Acyclic molecules were grouped by canonical SMILES. Groups were assigned to training, validation and test sets at an approximate ratio of 80:10:10. The allocation was designed to balance spectral modalities, data sources and measurement metadata, including phase, preparation and matrix availability. Detailed filters, duplicate-selection criteria and split counts are provided in Supplementary Section~\ref{app:experimental-curation} and Table~\ref{tab:dataset}.

\subsection{Equivariant harmonic-spectrum construction}\label{sec:methods-models}

\paragraph{Equivariant property prediction.}

We employ EquiformerV3\cite{equiformerv3} to encode atomic identities and three-dimensional coordinates using a radius graph with a 5~\angstrom\ cutoff. The production models comprise six interaction layers, 128 channels and a maximum angular-momentum degree of six. Property-specific equivariant heads predict tensor components in irreducible representations, which are subsequently converted to Cartesian form. Diagonal Hessian blocks are predicted from node features and off-diagonal blocks from directed edge features, while separate models predict dipole and polarizability derivatives. This approach avoids coordinate differentiation of learned energy and response functions.

\paragraph{Training objectives.}

The direct spectroscopic-tensor models are trained using smooth-L1 losses between predicted and reference quantities in their corresponding equivariant representations. For the alternative energy-gradient force model, energy and force predictions are jointly optimized using

\begin{equation}
\mathcal L_{\mathrm{EF}}
=
\mathcal L_{\mathrm{smoothL1}}
\left(\frac{E_\theta}{N},\frac{E_{\mathrm{ref}}}{N}\right)
+
\lambda\,\mathcal L_{\mathrm{smoothL1}}
\left(-\nabla_{\mathbf R}E_\theta,\mathbf F_{\mathrm{ref}}\right),
\end{equation}

where $N$ denotes the number of atoms. The energy loss is normalized per atom, while force errors are evaluated on their original component scale. Models are optimized using AdamW with cosine learning-rate decay, and checkpoints are selected by validation loss. The energy-gradient force objective uses $\lambda=80$. Run-specific hyperparameters and implementation details are specified in the accompanying code. Dataset populations and recorded training costs are provided in Supplementary Tables~\ref{tab:quantumdata} and \ref{tab:fig4-training-costs}.

\paragraph{Conformer generation and geometry optimization.}

Initial conformers are generated using RDKit ETKDGv3\cite{rdkit,wang2020etkdg} and refined with MMFF94 or UFF. Geometry optimization then uses directly predicted atomic forces with the FIRE algorithm\cite{bitzek2006fire}, supplemented by LBFGS refinement and alternative-conformer rescue when needed. The maximum-force convergence threshold is set to 0.02~eV/\angstrom.

The production workflow employs independently trained direct-force and energy models. An alternative energy-gradient approach, which computes forces as $\mathbf F=-\nabla_{\mathbf R}E_\theta$, serves as a fallback when direct-force relaxation fails.

\paragraph{Normal-mode analysis and spectral rendering.}

Harmonic-spectrum construction follows the normal-mode analysis procedure of DetaNet\cite{detanet}. The predicted diagonal and off-diagonal Hessian blocks are assembled into the Cartesian Hessian $H$. Using the diagonal atomic mass matrix $M$, we obtain

\begin{equation}
H_{\mathrm{sym}}=\tfrac12(H+H^{\mathsf T}),\qquad
D=M^{-1/2}H_{\mathrm{sym}}M^{-1/2},\qquad
D u_k=\lambda_k u_k.
\end{equation}

Vibrational frequencies are obtained from the eigenvalues after unit conversion. Cartesian dipole and polarizability derivatives are projected onto the mass-weighted normal modes to obtain normal-coordinate response derivatives. IR intensities depend on squared dipole derivatives, while Raman activities are determined by the isotropic and anisotropic components of the polarizability derivatives. Raman intensity conversion additionally accounts for excitation frequency and thermal populations\cite{wilson1980}.

Modes with imaginary frequencies are excluded from spectral rendering. The application renderer follows the DetaNet convention, using a fixed frequency scaling factor of 0.965 and Lorentzian broadening. Unscaled fixed-geometry fidelity tests and rigid-body-projected stability diagnostics use separate conventions, detailed in Supplementary Section~\ref{app:physical-evaluation}.

\subsection{HarmoNet for theory-to-experiment mapping}\label{sec:methods-harmonet}

\paragraph{Architecture and conditioning.}

HarmoNet learns the transformation from harmonic to experimental spectra by integrating molecular topology and measurement metadata. Each 1,024-point harmonic spectrum is divided into 64 non-overlapping patches, embedded with learned frequency positions and processed by six Transformer blocks.

Modality and metadata embeddings are combined into a conditioning vector $c$, which modulates spectral features through adaptive layer normalization\cite{peebles2023dit},

\begin{equation}
\mathrm{AdaLN}(z,c)=[1+\gamma(c)]\odot\mathrm{LN}(z)+\beta(c).
\end{equation}

Each Transformer block combines spectral self-attention, gated molecular cross-attention and conditioned feed-forward layers. The molecular graph encoder follows SpecGNN\cite{specgnn}, using three GINE layers\cite{hu2020pretraining} to encode atom and bond features. Explicit hydrogen atoms participate in message passing, while only heavy-atom embeddings are retained for cross-attention. A final adaptive normalization layer and linear patch decoder reconstruct the 1,024-point experimental spectrum, with negative predictions clamped to zero.

HarmoNet is trained using a masked mean-squared error (MSE) loss,

\begin{equation}
\mathcal L_{\mathrm{MSE}}=
\frac{1}{B}\sum_{i=1}^{B}
\frac{\sum_{p=1}^{1024}m_{ip}(\hat y_{ip}-y_{ip})^2}
{\max(1,\sum_{p=1}^{1024}m_{ip})},
\end{equation}

where $B$ denotes the batch size and $m_{ip}$ indicates whether a target spectral point is measured and finite. Each record is normalized by its number of valid points before batch averaging.

For application-level predictions, we add a seam loss (SEAM) that matches first differences between predicted and experimental spectra at patch boundaries to reduce discontinuities. Let $\mathcal B$ denote the left indices of patch boundaries, with $v_{ib}=m_{ib}m_{i,b+1}$ and $\Delta_b y_i=y_{i,b+1}-y_{ib}$. The seam loss is defined as

\begin{equation}
\mathcal L_{\mathrm{seam}}=
\frac{1}{B}\sum_{i=1}^{B}
\frac{\sum_{b\in\mathcal B}v_{ib}
(\Delta_b\hat y_i-\Delta_b y_i)^2}
{\max(1,\sum_{b\in\mathcal B}v_{ib})},
\qquad
\mathcal L=\mathcal L_{\mathrm{MSE}}+0.1\mathcal L_{\mathrm{seam}}.
\end{equation}

The main information-source ablations use the MSE objective alone. HarmoNet is optimized using Adam with cosine learning-rate decay, and checkpoints are selected by validation cosine similarity. Metadata conditioning includes values and their availability/inference status; source identifiers are excluded. Input-removal controls retain the matched architecture where applicable, while graph removal also removes cross-attention. SpecGNN uses its three-layer GINE architecture with the same masked MSE objective; Chemprop-IR retains its directed message-passing architecture and SID objective\cite{specgnn,chemprop}. Peptide-specific adaptation and checkpoint selection are detailed in Supplementary Section~\ref{app:peptides}.

\subsection{Spectrum-to-structure modeling}
\label{sec:methods-inverse}

\paragraph{Pseudo-experimental pretraining.}

HARMONY generates pseudo-experimental spectra for the VB-Mols training and validation molecules, replacing DFT spectra in Vib2Mol pretraining while retaining its native architecture, experimental fine-tuning and inference workflow\cite{vib2mol}. Pseudo-source molecules exclude the union of legacy NIST/SDBS validation and test connectivities. Metadata tuples are sampled separately for IR and Raman from frequently observed patterns in the joint experimental training records, with probabilities proportional to their frequencies and all-unknown tuples excluded. The spectra and molecular formulas are supplied to Vib2Mol; the sampled metadata are used by HARMONY to generate the spectra. Vib2Mol retains its native matching and autoregressive objectives and 1,000-epoch training schedule\cite{vib2mol}.

\paragraph{Spectral reranking through forward prediction.}

We replace Vib2Mol's native matching-based reranking with forward spectral verification using HARMONY. Given a formula-filtered pool of candidate structures, HARMONY predicts the experimental spectrum of each candidate under the measurement conditions associated with the input spectrum. IR and Raman spectra are predicted separately using their respective metadata, with the experimental conditions held fixed across candidates. Predicted spectra are clipped to non-negative values and peak-normalized before comparison.

For candidate molecule $M_i$ and spectral modality $t$, the verification score is calculated as the cosine similarity between the predicted spectrum $\hat{\mathbf{s}}_{i,t}$ and the measured spectrum $\mathbf{s}_t$ on the same 1,024-point grid,

\begin{equation}
c_{i,t}=
\frac{\hat{\mathbf{s}}_{i,t}^{\mathsf T}\mathbf{s}_t}
{\lVert\hat{\mathbf{s}}_{i,t}\rVert_2
 \lVert\mathbf{s}_t\rVert_2}.
\end{equation}

When multiple modalities are available, their similarities are averaged to obtain the final ranking score,

\begin{equation}
R_i=\frac{1}{|\mathcal T|}\sum_{t\in\mathcal T}c_{i,t},
\end{equation}

where $\mathcal T$ denotes the set of available experimental modalities. Joint IR and Raman inference therefore assigns equal weight to both modalities, while single-modality inference uses the corresponding similarity alone.

Candidate structures are ranked in descending order of $R_i$, without incorporating generation probabilities or native matching scores.

For forward verification, HarmoNet and SpecGNN are pretrained on a HarmoSpec-derived base excluding legacy NIST/SDBS validation and test connectivities, then fine-tuned on the joint legacy training data. Test identities are excluded from validation. Fine-tuning uses a reduced learning rate, and the verifier checkpoint is selected by validation cosine similarity. Measurement metadata are recovered by molecular identity and source-record matching, using spectral similarity to resolve ambiguous matches.

\paragraph{Candidate-pool expansion.}

We combine the validation-selected and final-epoch Vib2Mol checkpoints from the same training run, deduplicating candidates by InChIKey. Each checkpoint uses beam size 10, temperature 3.5 and up to five formula-valid candidates. Lists are merged in round-robin order, yielding up to ten distinct candidates. SpecGNN and single-pool HARMONY reranking use identical candidate files.

\subsection{Agent-assisted structure elucidation}
\label{sec:methods-agent}

We constructed natural-language structure-elucidation questions from the NIST and SDBS test sets. NIST questions provide numerical IR spectra, while SDBS questions provide Raman, IR or combined IR--Raman spectra; all settings include a molecular formula and available measurement metadata. For NIST, the same retrospectively inferred measurement metadata were supplied to Codex alone, the Vib2Mol--HARMONY Agent and the fixed workflow; unresolved conditions remained unknown. Molecular names, reference structures and evaluator-only identities were withheld. Each question was processed in a fresh Codex CLI session using \texttt{gpt-6-astra} at \texttt{xhigh} reasoning effort, with web search and external reference retrieval disabled.

Scientific capabilities were provided through local Model Context Protocol (MCP) tools, including Vib2Mol candidate generation, molecular structure and formula validation, HARMONY spectral prediction, spectral comparison and peak extraction. Experimental conditions were handled separately for IR and Raman, with missing metadata distinguished from explicitly absent conditions.

The guided search begins with candidate generation using Vib2Mol. The agent can expand the candidate pool by using additional checkpoints, increasing beam sizes, adjusting generation settings or independently proposing molecular structures. Search guidance provides advisory limits of three generation attempts and 20 distinct candidates for spectral verification; these are agent-side instructions rather than server-enforced limits. When initial candidates are insufficient, the agent analyzes spectral features and proposes alternatives, checks their molecular formulas and validity, and submits eligible structures for forward verification. Candidate origins and tool calls are recorded. A tentative first choice is requested when plausible candidates remain; abstention is permitted when no defensible candidate is found.

\paragraph{Fixed-workflow and agent comparisons.}

The fixed workflow uses the merged candidate pool and HARMONY reranking described above. The Vib2Mol--HARMONY Agent uses the same generator pair and verifier, with adaptive search and additional structure proposals. Codex alone receives the same questions without access to scientific tools. An example natural-language question is provided in Supplementary Section~\ref{app:agent-outcomes}; paired outcomes and fixed-pool coverage are reported in Supplementary Table~\ref{tab:agent-candidate-analysis}.

\subsection{Evaluation metrics and computational cost}\label{sec:methods-evaluation}

\paragraph{Spectral prediction metrics.}

Predicted and experimental spectra are evaluated on the common 1,024-point grid using only valid measured regions. Negative intensities are clipped to zero, and both spectra are independently min--max normalized over the valid region. The primary metric is cosine similarity,

\begin{equation}
C_i=
\frac{\sum_{p\in\mathcal V_i}a_{ip}b_{ip}}
{\sqrt{\sum_{p\in\mathcal V_i}a_{ip}^{2}}
 \sqrt{\sum_{p\in\mathcal V_i}b_{ip}^{2}}},
\end{equation}

where $\mathcal V_i$ denotes the valid spectral points and $\mathbf a_i$ and $\mathbf b_i$ represent the normalized predicted and experimental spectra, respectively. Mean similarities are calculated across experimental records within each modality. Predicted spectra use Savitzky--Golay smoothing with a 14-point window and cubic polynomial by default. The input and fusion comparisons in Fig.~\ref{fig:ablation}A and Supplementary Tables~\ref{tab:ablations} and \ref{tab:raman-ablation-trajectory} apply this same postprocessing to every model, including SpecGNN.

Regional input ablations use the same test records and five windows: 400--1000, 1000--1500, 1500--1800, 1800--2500 and 2500--4000~\wn. Each window requires at least 90\% valid coverage, a target peak of at least 0.05 after full-spectrum normalization and nonzero cosine norms. Window-specific cosine losses retain the full-spectrum normalization. Differences between the complete and separately trained input-removal models are averaged within molecular connectivity groups, including both modalities where available, and then equally across groups. Pointwise 95\% confidence intervals use 5,000 paired molecule-bootstrap resamples.

\paragraph{Molecular structure inference metrics.}

Structure generation is evaluated using Top-$k$ recall, defined as the fraction of test records for which the reference structure appears among the first $k$ ranked candidates. Invalid predictions, empty candidate pools and unanswered questions are counted as incorrect. Candidate-pool oracle recall measures reference coverage independently of ranking performance. Retrieval Recall@$k$ is reported separately because retrieval assumes a predefined molecular library, whereas generation must produce the correct structure.

Native Vib2Mol results follow the original molecular-identity evaluation protocol. For fixed-workflow and agent comparisons, molecular identities are matched using standard InChIKeys, with unspecified reference E/Z stereochemistry excluded from the corresponding comparison. Reference-specified stereochemistry and other molecular attributes remain enforced. All NIST comparisons use the complete 1,215-record test set. Forward rerankers produced no predictions for 17 multicomponent records; these are counted as incorrect rather than excluded. Codex and agent scores include their actual answers to all 1,215 questions. Each SDBS setting uses all 282 records. The literature comparison in Fig.~\ref{fig:inverse-enablement}E uses the values reported under each original study's evaluation protocol.

\paragraph{Computational latency.}

Accuracy--efficiency comparisons use a fixed panel of 45 molecules with experimental gas-phase IR spectra, selected to cover a range of molecular sizes and chemical structures. All primary workflows are evaluated on the same molecules; the additional DetaNet-MD route contributes 44 quality-controlled trajectories and is reported separately.

Learned-workflow latency is measured using serial, single-molecule execution on an NVIDIA H100 GPU after model initialization and warmup. For HARMONY, timing includes conformer generation, force-field refinement, geometry optimization, tensor prediction, harmonic-spectrum construction and HarmoNet inference. Model loading, startup overhead and dataset I/O are excluded. Molecular-dynamics timing includes trajectory sampling and spectral analysis. Both MD workflows run at 300~K: MACE4IR uses 5~ps of equilibration followed by 50~ps of production, while DetaNet-MD uses 10~ps and 40~ps, respectively. MACE4IR frequencies are scaled by 0.965; DetaNet-MD uses a smooth transition from unscaled low-frequency modes to the same scaling at high frequencies, selected on the comparison panel. Panel selection and workflow timing are detailed in Supplementary Section~\ref{app:workflow-timing}.

\section*{Acknowledgements}

This work was supported by the National Natural Science Foundation of China (Grant Nos. 22595412, 22474117, 92570205 and 92461313), the Fundamental Research Funds for the Central Universities (Grant No. 20720250080), the Chuangqi Program of the Shanghai Innovation Institute, and the Shanghai Municipal Education Commission.


\section*{Competing interests}
The authors declare no competing interests.

\FloatBarrier
\bibliography{references}
\clearpage
\begingroup
\setcounter{section}{0}
\setcounter{subsection}{0}
\setcounter{figure}{0}
\setcounter{table}{0}
\setcounter{equation}{0}
\setcounter{page}{1}
\renewcommand{\thesection}{S\arabic{section}}
\renewcommand{\thefigure}{S\arabic{figure}}
\renewcommand{\thetable}{S\arabic{table}}
\renewcommand{\theequation}{S\arabic{equation}}
\renewcommand{\thepage}{S\arabic{page}}
\renewcommand{\theHsection}{supplement.\arabic{section}}
\renewcommand{\theHfigure}{supplement.\arabic{figure}}
\renewcommand{\theHtable}{supplement.\arabic{table}}
\renewcommand{\theHequation}{supplement.\arabic{equation}}
\pagestyle{fancy}
\fancyhf{}
\fancyfoot[C]{\thepage}
\renewcommand{\headrulewidth}{0pt}
\begin{center}
{\Large\bfseries Supplementary Information\par}
\end{center}

\FloatBarrier
\section{Dataset composition, curation and splits}\label{app:experimental-curation}\label{app:training-records}

Spectral preprocessing follows Methods. Curation retains single-component molecules with validated identities and rejects identical experimental targets assigned to conflicting molecular identities. Near-duplicates are compared within groups sharing molecular identity, modality and measurement metadata. A match requires cosine similarity of at least 0.9999 and RMSE of at most 0.003 over a shared valid region containing at least 512 points and covering at least 98\% of the smaller validity mask. Representatives are selected by signal quality, prioritizing signal-to-noise ratio, coverage and trace confidence. Final record and molecule counts are given in Table~\ref{tab:dataset}.

When multiple DFT calculations are available for a molecule, the representative maximizes mean pairwise cosine similarity across harmonic IR and Raman spectra. The scaffold-based split and metadata fields follow Methods.

\begin{table}[htbp]
\caption{HarmoSpec experimental records and molecular coverage. Each record contains one measured modality. Distinct measurements of the same molecule may be retained.}\label{tab:dataset}
\centering
\begin{tabular}{lrrrr}
\toprule
Split & Molecules & IR & Raman & Total records\\
\midrule
Training & 15,283 & 26,738 & 2,018 & 28,756\\
Validation & 1,863 & 3,342 & 252 & 3,594\\
Test & 1,921 & 3,343 & 252 & 3,595\\
\midrule
Total & 19,067 & 33,423 & 2,522 & 35,945\\
\bottomrule
\end{tabular}
\end{table}

\begin{table}[ht]
\caption{Quantum-reference record counts. QMe14S-SP includes multiple configurations per molecule.}\label{tab:quantumdata}
\centering
\begin{tabular}{lrrr}
\toprule
Dataset & Training & Validation & Test\\
\midrule
QM9S & 116,789 & 6,537 & 6,491\\
QMe14S-opt & 167,425 & 9,371 & 9,306\\
QMe14S-sp & 485,574 & 27,176 & 26,959\\
\bottomrule
\end{tabular}
\end{table}

\FloatBarrier

\section{Gas-phase panel selection and timing}\label{app:workflow-timing}

The benchmark comprises 45 gas-phase IR molecules from the HarmoSpec test set, selected from neutral, closed-shell, single-component molecules containing 4--14 heavy atoms with at least 95\% spectral coverage. Each molecule has a validated Gaussian optimization and frequency calculation without imaginary frequencies. Molecules were selected across six heavy-atom strata using RDKit MaxMin diversity sampling based on Morgan fingerprints. All workflows share the same fixed panel, with learned relaxation and MD initialized from ETKDGv3 conformers refined using MMFF94 or UFF.

Gaussian latency includes geometry optimization and frequency calculations, performed primarily on 64 CPU cores, with a mean of 111.96~s per molecule. Learned harmonic workflows were benchmarked on an NVIDIA H100 GPU using sequential, uncached SMILES-to-spectrum inference. Timing covers conformer generation, force-field refinement, geometry relaxation, tensor prediction, normal-mode analysis and final spectral prediction, including data preparation and device transfers. GPU operations were synchronized, while model loading and initial warmup were excluded. End-to-end neural predictors were evaluated at batch size one, using 50 warmup calls followed by 100 timed calls per molecule.

\section{Complementary spectral, graph and metadata information}\label{app:results}

\FloatBarrier
\subsection{Input and fusion controls}

Tables~\ref{tab:ablations} and \ref{tab:raman-ablation-trajectory} report IR and Raman input and fusion ablations with SG(14,3) postprocessing applied to every model. These controls use DFT-derived harmonic inputs; the fully learned production route is evaluated separately in the direct-prediction experiments. Global graph conditioning is an intermediate fusion control; the final HarmoNet uses all-heavy-node cross-attention.

\begin{table}[ht]
\caption{IR input and fusion ablations after SG(14,3) postprocessing. Graph-only controls retain the HarmoNet decoder architecture.}\label{tab:ablations}
\centering
\begin{tabular}{lr}
\toprule
Configuration & Mean cosine\\
\midrule
Harmonic only & 0.836\\
Harmonic + phase & 0.865\\
Harmonic + phase + preparation & 0.894\\
Harmonic + phase + preparation + matrix & 0.895\\
All inputs, global graph conditioning & 0.908\\
All inputs, all-heavy-node cross-attention (final) & 0.911\\
\midrule
Matched graph only & 0.830\\
Matched graph + harmonic, no metadata & 0.849\\
Matched graph + metadata, no harmonic & 0.888\\
\midrule
Pooled graph tokens: $K=1$ & 0.909\\
Pooled graph tokens: $K=2$ & 0.910\\
Pooled graph tokens: $K=4$ & 0.909\\
Pooled graph tokens: $K=8$ & 0.910\\
Pooled graph tokens: $K=16$ & 0.910\\
\bottomrule
\end{tabular}
\end{table}

\begin{table}[!htbp]
\centering
\caption{Raman input and fusion ablations after SG(14,3) postprocessing.}
\label{tab:raman-ablation-trajectory}
\begin{tabular}{lc}
\toprule
Model / information & Mean cosine \\
\midrule
SpecGNN (graph only) & 0.874 \\
HarmoNet (graph only, matched architecture) & 0.730 \\
HarmoNet (harmonic only) & 0.889 \\
HarmoNet (graph + metadata) & 0.886 \\
HarmoNet (harmonic + phase) & 0.894 \\
HarmoNet (harmonic + phase + preparation) & 0.896 \\
HarmoNet (harmonic + phase + preparation + matrix) & 0.897 \\
HarmoNet (all inputs, global graph fusion) & 0.904 \\
HarmoNet (all inputs, all-heavy cross-attention) & 0.905 \\
\bottomrule
\end{tabular}
\end{table}
\FloatBarrier

\subsection{Fingerprint-region retraining}\label{app:restricted-window}

To assess prediction in the fingerprint region, we retrained the complete HarmoNet and three input-ablation models on 400--2000~\wn, retaining the original training settings and dataset split while restricting both harmonic inputs and targets to this range. When evaluated over 400--2000~\wn, restricted-range and full-range training yield IR cosines of 0.901 and 0.902, and Raman cosines of 0.832 and 0.833, respectively. Figure~\ref{fig:restricted-window-ablation} compares input-ablation losses over each model's training range. All three information sources remain beneficial, with metadata providing the largest contribution to IR prediction and harmonic information contributing most strongly to Raman prediction in both settings.

\begin{figure}[htbp]
\centering
\includegraphics[width=.85\linewidth]{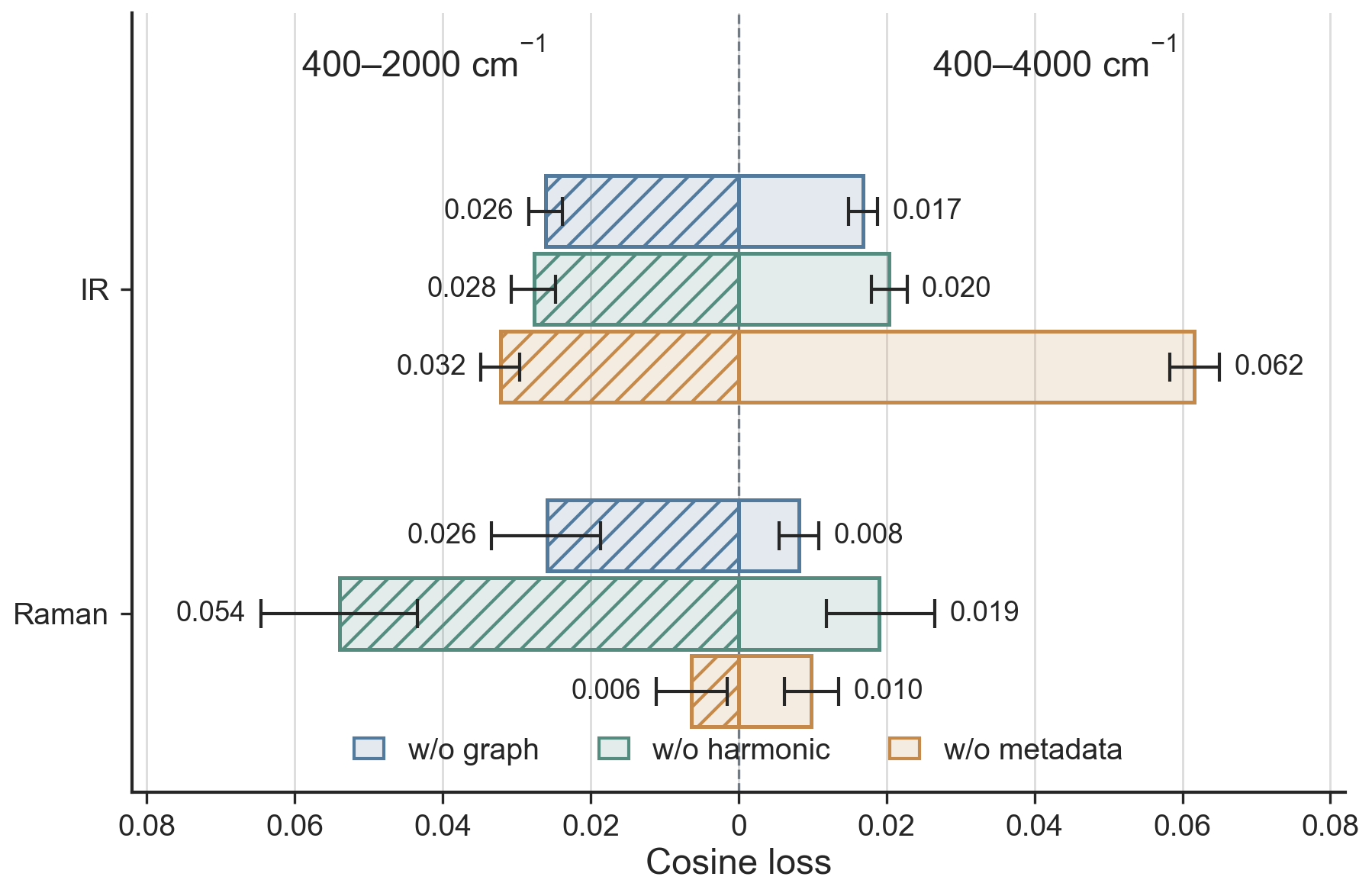}
\caption{Input-ablation losses for models trained and evaluated on 400--2000~\wn\ (left, hatched) or 400--4000~\wn\ (right, unhatched). Bar lengths show molecule-weighted full-minus-ablated cosine for IR and Raman; leftward placement identifies the restricted window, with both sides reporting positive losses. Error bars show pointwise 95\% paired molecule-bootstrap intervals (5,000 resamples; seed 624). Colors match main-text Fig.~\mainfigref{fig:ablation}C.}\label{fig:restricted-window-ablation}
\end{figure}

\section{Metadata-resolved prediction performance}\label{app:metadata-controls}

Figure~\ref{fig:metadata-tree} summarizes performance across the recorded phase, preparation and matrix combinations.

\begin{figure}[p]
\centering
\includegraphics[width=\linewidth,height=0.85\textheight,keepaspectratio]{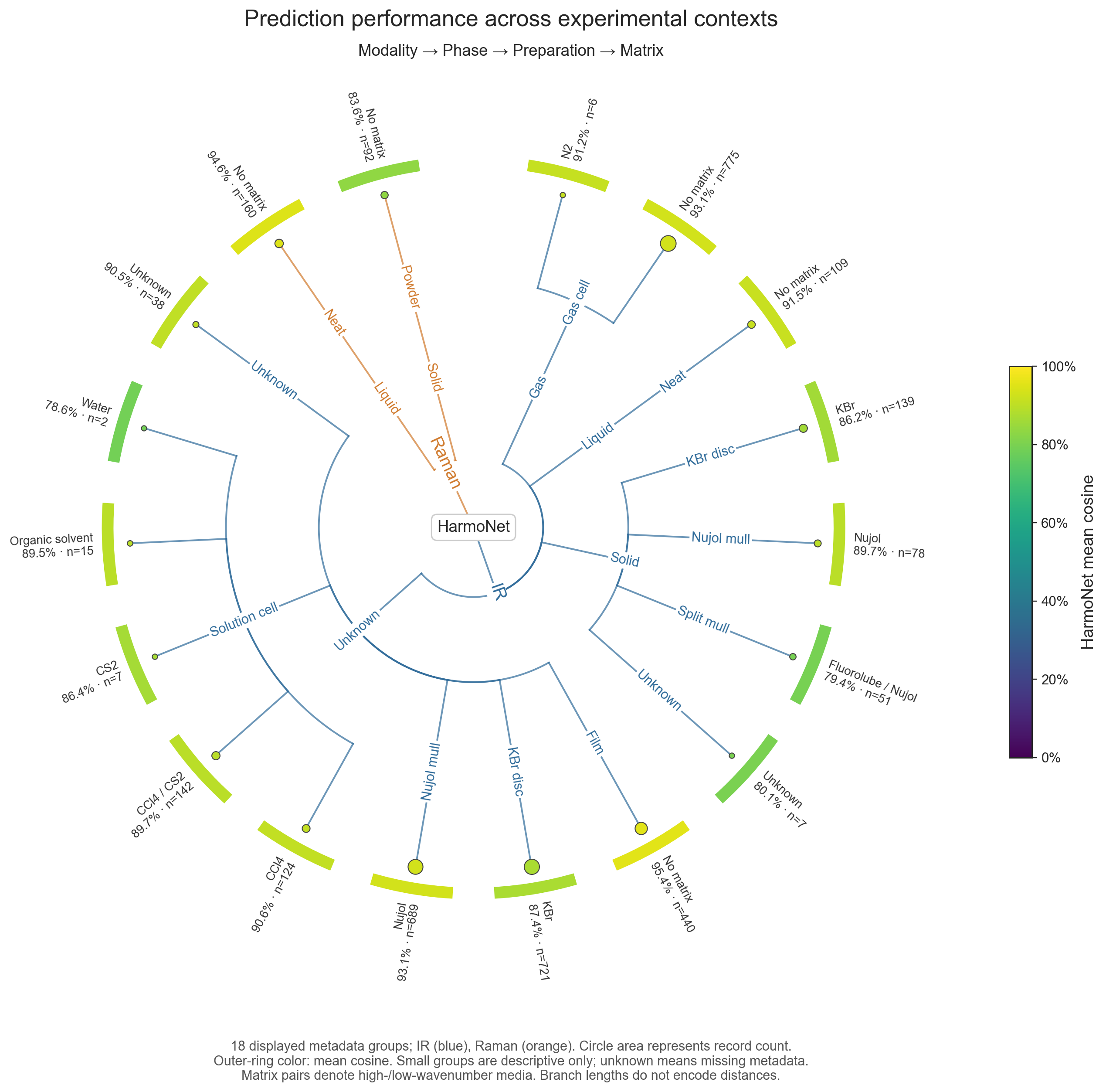}
\caption{Performance across the 18 recorded metadata combinations in the 3,595-record test set. The tree follows modality, phase, preparation and matrix. Colors encode record-weighted mean cosine; circle areas and labels report record counts. Unknown denotes missing metadata; paired matrix names identify high- and low-wavenumber media.}\label{fig:metadata-tree}
\end{figure}

\FloatBarrier
\section{Direct prediction and complete-workflow evaluation}\label{app:direct-prediction}

\FloatBarrier
\subsection{Property accuracy and computational cost}\label{app:tensor-results}

Direct models predict forces and spectroscopic tensors through equivariant output heads; gradient models obtain coordinate derivatives of predicted energy, dipole or polarizability. Tables~\ref{tab:fig4-training-costs} and \ref{tab:tensor-grad-direct} report the recorded GPU-hour costs and test errors. Training costs sum runtime across participating GPUs. Tables~\ref{tab:qm9s-properties} and \ref{tab:qme14s-properties} provide multi-property MAEs on QM9S and QMe14S-SP, including DetaNet and XPaiNN comparisons\cite{detanet,xpainn}. Production energy, force and tensor models are trained on QMe14S and frozen during HarmoNet training and inference.

\begin{table}[!htbp]
\centering
\small
\caption{Dipole-derivative training costs on QM9S, reported in summed GPU hours.}\label{tab:fig4-training-costs}
\begin{tabular}{lrrr}
\toprule
Configuration & Grad (GPU h) & Direct (GPU h) & Grad/Direct \\
\midrule
DetaNet & 16.14 & 3.54 & 4.56 \\
L3-C32-D3 & 40.70 & 5.97 & 6.82 \\
L3-C64-D3 & 61.44 & 8.58 & 7.16 \\
L3-C128-D3 & 132.48 & 14.61 & 9.07 \\
L3-C128-D4 & 256.70 & 18.96 & 13.54 \\
L3-C128-D5 & 296.45 & 23.34 & 12.70 \\
L3-C128-D6 & 341.76 & 27.94 & 12.23 \\
L4-C128-D6 & 385.15 & 33.24 & 11.59 \\
L5-C128-D6 & 421.25 & 43.04 & 9.79 \\
L6-C128-D6 & 494.40 & 50.54 & 9.78 \\
\bottomrule
\end{tabular}
\end{table}

\begin{table}[htbp]
\caption{QM9S gradient/direct tensor-prediction MAEs and training costs (GPU hours).}
\label{tab:tensor-grad-direct}
\centering\small
\setlength{\tabcolsep}{3pt}
\begin{tabular}{llrrrr}
\toprule
& & \multicolumn{2}{c}{Test MAE} & \multicolumn{2}{c}{Training (GPU h)}\\
Model & Property & Grad & Direct & Grad & Direct\\
\midrule
DetaNet & Hessian & 0.0548 & 0.1385 & 75.96 & 7.01\\
DetaNet & Polarizability derivative & 0.1594 & 0.4698 & 28.15 & 3.55\\
EqV3-L6-C128-D6 & Hessian & 0.0135 & 0.0054 & 3,645.17 & 104.35\\
EqV3-L6-C128-D6 & Polarizability derivative & 0.0310 & 0.0391 & 1,329.68 & 50.59\\
\bottomrule
\end{tabular}
\par\smallskip
\begin{minipage}{\linewidth}\footnotesize
Hessian costs sum separately trained diagonal and off-diagonal models; the EqV3 gradient cost includes eight DDP GPUs.
Hessian MAEs refer to the complete tensor. Errors retain the property-specific reporting scales.
\end{minipage}
\end{table}

\begin{table}[!htbp]
\centering\small
\caption{Multi-property test MAEs on QM9S, using the reporting scale of each property. A dash indicates an unavailable result.}
\label{tab:qm9s-properties}
\begin{tabular}{lrrrr}
\toprule
Property & DetaNet & EnviroDetaNet & XPaiNN & EqV3-direct\\
\midrule
Polarizability derivative & 0.2013 & 0.1096 & --- & 0.0391\\
Dipole & 0.0087 & 0.0067 & 0.0058 & 0.0039\\
Polarizability & 0.1374 & 0.0657 & 0.0543 & 0.0411\\
Hyperpolarizability & 2.0369 & 1.5760 & 1.3947 & 1.6543\\
Quadrupole & 0.3834 & 0.2682 & 0.2208 & 0.1307\\
Octupole & 6.2942 & 5.6794 & 4.1583 & 4.5230\\
\bottomrule
\end{tabular}
\end{table}

\begin{table}[!htbp]
\centering\small
\caption{Multi-property test MAEs on QMe14S-SP.}
\label{tab:qme14s-properties}
\begin{tabular}{lrr}
\toprule
Property / unit & DetaNet & EqV3\\
\midrule
Energy / eV & 0.0402 & 0.0218\\
Forces / eV\,\AA$^{-1}$ & 0.0362 & 0.0080\\
Dipole / D & 0.0285 & 0.0101\\
Polarizability / \AA$^3$ & 0.2611 & 0.1081\\
\bottomrule
\end{tabular}
\end{table}

\FloatBarrier
\subsection{Spectroscopic fidelity and physical diagnostics}\label{app:physical-evaluation}

Figure~\ref{fig:harmonic-fidelity} and Tables~\ref{tab:spectral-fidelity} and \ref{tab:harmonic-spectrum-fidelity} report frequency errors and reconstructed harmonic spectra at fixed DFT geometries. Tables~\ref{tab:physical-diagnostics}, \ref{tab:internal-mode-stability} and \ref{tab:tensor-equivariance} report tensor consistency, projected vibrational stability and rotation equivariance, respectively.

\begin{figure}[!htbp]
\centering
\includegraphics[width=\linewidth]{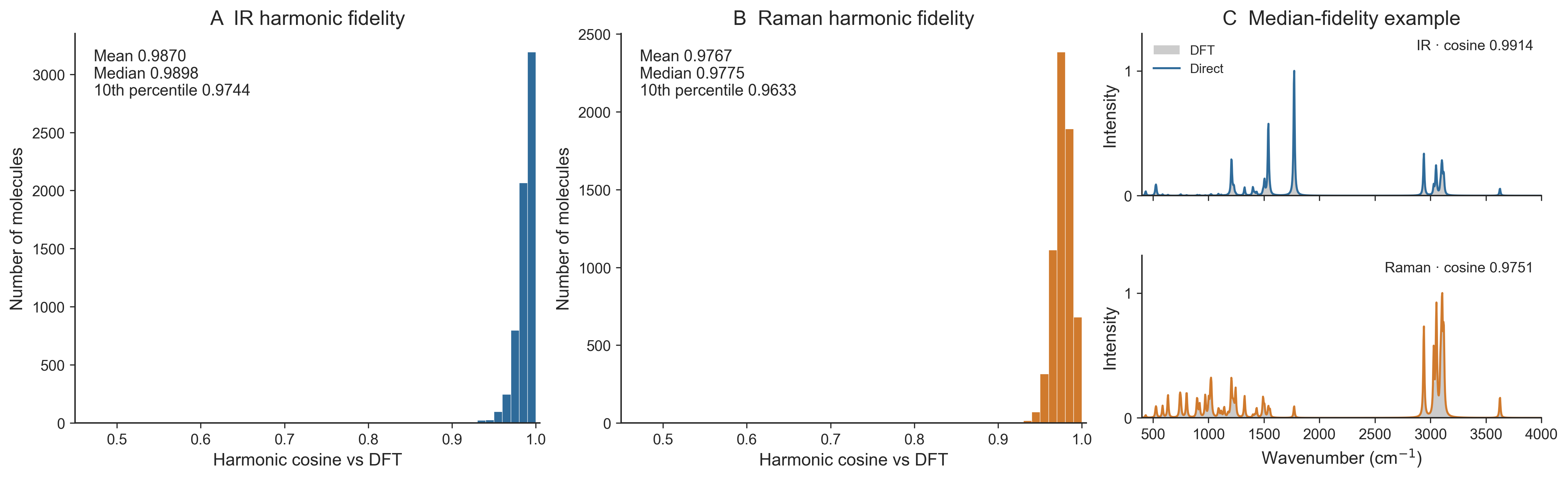}
\caption{Harmonic-spectrum fidelity at DFT geometries for 6,491 QM9S test molecules. (A,B) IR and Raman cosine distributions. (C) A molecule nearest the median of mean IR/Raman cosine. Rendering settings are given in Table~\ref{tab:harmonic-spectrum-fidelity}.}\label{fig:harmonic-fidelity}
\end{figure}

\begin{table}[!htbp]
\centering\small
\caption{Per-molecule vibrational-frequency errors at DFT geometries (\wn).}\label{tab:spectral-fidelity}
\begin{tabular}{lrlrrrr}
\toprule
Dataset & $n$ & Error & Mean & Median & 95th percentile & Maximum\\
\midrule
QM9S & 6,491 & MAE & 1.305 & 1.278 & 1.733 & 21.945\\
QM9S & 6,491 & RMSE & 2.002 & 1.881 & 3.043 & 27.431\\
QMe14S-opt & 9,306 & MAE & 1.606 & 1.387 & 2.554 & 94.030\\
QMe14S-opt & 9,306 & RMSE & 2.615 & 2.206 & 4.830 & 116.103\\
\bottomrule
\end{tabular}
\par\smallskip
\begin{minipage}{\linewidth}\footnotesize
Frequencies use scale one and are paired by sorted rank after excluding five rigid-body modes for linear molecules or six for nonlinear molecules. Frequency errors weight molecules equally.
\end{minipage}
\end{table}

\begin{table}[!htbp]
\centering\small
\caption{Harmonic-spectrum cosine similarities against DFT references for 6,491 QM9S test molecules at the same geometries.}\label{tab:harmonic-spectrum-fidelity}
\begin{tabular}{lrrrrr}
\toprule
Observable & Mean & SD & 5th percentile & Median & 95th percentile\\
\midrule
IR intensity & 0.987 & 0.016 & 0.967 & 0.990 & 0.999\\
Raman activity & 0.970 & 0.015 & 0.946 & 0.972 & 0.989\\
Raman intensity & 0.977 & 0.012 & 0.958 & 0.977 & 0.994\\
\bottomrule
\end{tabular}
\par\smallskip
\begin{minipage}{\linewidth}\footnotesize
Spectra use a Lorentzian FWHM of 12~\wn; Raman intensity is rendered at 532~nm. SD denotes variation across molecules.
\end{minipage}
\end{table}

\begin{table}[!htbp]
\centering\small
\caption{Tensor consistency and vibrational density-of-states (VDOS) fidelity at DFT geometries.}\label{tab:physical-diagnostics}
\begin{tabular}{lrr}
\toprule
Metric & QM9S & QMe14S-opt\\
\midrule
Off-diagonal reciprocity residual & 2.77\% & 3.85\%\\
Translational sum-rule residual & 0.205\% & 0.247\%\\
VDOS cosine against DFT & 0.999 & 0.997\\
\bottomrule
\end{tabular}
\par\smallskip
\begin{minipage}{\linewidth}\footnotesize
Off-diagonal reciprocity is measured before Hessian symmetrization.
\end{minipage}
\end{table}

\begin{table}[!htbp]
\centering\small
\caption{Projected internal-mode stability of direct and DFT Hessians.}\label{tab:internal-mode-stability}
\begin{tabular}{lrrrrr}
\toprule
 & & \multicolumn{4}{c}{Frequency threshold (\wn)}\\
\cmidrule(lr){3-6}
Dataset / Hessian & $n$ & $0$ & $-20$ & $-50$ & $-100$\\
\midrule
QM9S / direct & 6,491 & 96.13 & 97.47 & 99.35 & 100.00\\
QM9S / DFT & 6,491 & 99.38 & 99.86 & 99.98 & 100.00\\
QMe14S-opt / direct & 9,306 & 90.66 & 93.62 & 98.59 & 99.90\\
QMe14S-opt / DFT & 9,306 & 98.94 & 99.59 & 99.84 & 99.88\\
HarmoSpec relaxed / direct & 1,921 & 49.61 & 58.82 & 87.09 & 98.80\\
\bottomrule
\end{tabular}
\par\smallskip
\begin{minipage}{\linewidth}\footnotesize
Values are percentages of molecules whose lowest internal-mode frequency is at or above each signed threshold after rigid-body projection. Quantum benchmarks use DFT geometries; the HarmoSpec row uses learned relaxed geometries.
\end{minipage}
\end{table}

\begin{table}[!htbp]
\centering\small
\caption{Numerical rotation equivariance of directly predicted spectroscopic tensors.}\label{tab:tensor-equivariance}
\begin{tabular}{lrr}
\toprule
Property & Mean & 95th percentile\\
\midrule
Diagonal Hessian & 0.00187 & 0.00292\\
Off-diagonal Hessian & 0.00262 & 0.00436\\
Dipole derivative & 0.00251 & 0.00487\\
Polarizability derivative & 0.00320 & 0.00487\\
\bottomrule
\end{tabular}
\par\smallskip
\begin{minipage}{\linewidth}\footnotesize
Values are relative Frobenius errors over 200 QMe14S-opt geometries and three rotations per geometry.
\end{minipage}
\end{table}

\FloatBarrier

\subsection{Direct-force optimization coverage}\label{app:direct-force-validation}

We evaluated direct-force optimization during construction of the fully learned HarmoSpec dataset. Across the training, validation and test splits, 19,052 of 19,067 molecules (99.92\%) reached a maximum predicted atomic-force norm below 0.02~eV\,\AA$^{-1}$ using the direct-force optimization pipeline, including its rescue steps. The remaining 15 training molecules converged through the energy-gradient fallback, giving complete coverage. All validation and test molecules used direct-force optimization.

\begin{table}[htbp]
\centering\small
\caption{Direct-force optimization coverage across all three HarmoSpec splits, including the direct pipeline's rescue steps. Convergence requires a maximum predicted atomic-force norm below 0.02~eV\,\AA$^{-1}$.}\label{tab:direct-force-validation}
\begin{tabular}{lr}
\toprule
Production optimization & Molecules\\
\midrule
Direct-force convergence & 19,052/19,067 (99.92\%)\\
Energy-gradient fallback & 15\\
Total convergence & 19,067/19,067 (100.00\%)\\
\bottomrule
\end{tabular}
\end{table}

\subsection{Progressive DFT replacement and cross-route evaluation}

The full-test workflow comparison in main-text Fig.~\mainfigref{fig:direct-harmonic}D covers 1,921 HarmoSpec molecules, with IR and Raman accuracy evaluated over 3,343 and 252 experimental records, respectively. DFT timing combines optimization and frequency-calculation logs, using 55 CPU cores for 1,067 molecules and 64 cores for 854 molecules. Learned-route latency sums separately measured warm, serial single-molecule components on an H100 GPU. These full-test measurements are distinct from the 45-molecule panel described in Section~\ref{app:workflow-timing}.

Figure~\ref{fig:cross-route-progressive} holds each HarmoNet checkpoint fixed while exchanging DFT, energy-gradient-relaxed and direct-force-relaxed harmonic inputs. Targets, masks and metadata remain unchanged. Both learned-relaxation routes use directly predicted spectroscopic tensors. Each row compares alternative harmonic inputs against the same checkpoint's original input route.

\begin{figure}[!htbp]
\centering
\includegraphics[width=\linewidth]{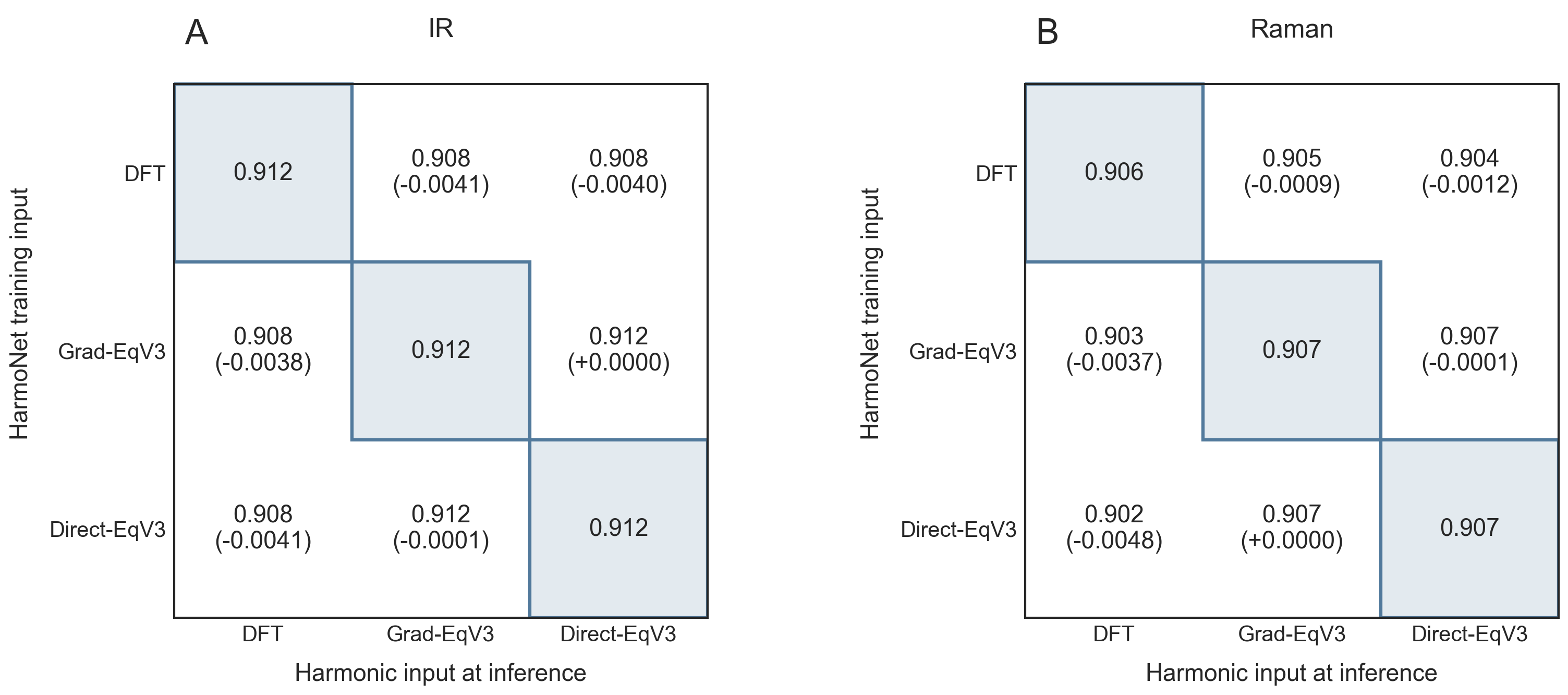}
\caption{Fixed-checkpoint cross-route robustness for (A) IR and (B) Raman. Values are mean experimental-spectrum cosine similarities, with training routes in rows and inference routes in columns. Highlighted diagonal cells identify matched routes; parentheses show changes relative to that row's matched entry. Grad and Direct refer to relaxation forces; both use direct spectroscopic tensors.}\label{fig:cross-route-progressive}
\end{figure}

\section{Natural-language benchmark example and paired outcomes}\label{app:agent-outcomes}

The following question (SDBS task \texttt{Qd1dcc4b8bab6}) illustrates the natural-language inputs used for structure elucidation. The two spectrum files accompany the question; the reference molecular identity is not supplied.

\begin{quote}\small
I have spectra of an unknown sample with the molecular formula C9H18O2. The IR spectrum is in \texttt{ir.txt}, and the Raman spectrum is in \texttt{raman.txt}. The IR spectrum was measured using a film sample, and the Raman spectrum using a neat liquid sample.

What is the most likely molecular structure? Please give your first choice as a SMILES string. If there are other plausible structures, include them as well (no more than five candidates in total), and explain which spectral features support your assessment. If the structure cannot be determined with confidence, explain why.

Both files have two columns: wavenumber (\wn) followed by nonnegative normalized intensity, not raw transmittance. Use the measurement conditions given for each spectrum separately, and do not assume any conditions that have not been specified.
\end{quote}

Table~\ref{tab:agent-candidate-analysis} reports paired outcomes conditional on reference coverage in the fixed candidate pool.

\begin{table}[!htbp]
\centering\small
\caption{Joint fixed-pool coverage and paired outcomes for NIST IR and SDBS IR--Raman structure elucidation, with molecular formulas provided.}\label{tab:agent-candidate-analysis}
\begin{tabular}{llrrr}
\toprule
Reference in fixed pool & Fixed workflow & Agent correct & Agent incorrect & Total\\
\midrule
\multicolumn{5}{l}{\textbf{NIST IR ($n=1{,}215$)}}\\
Present & Correct & 717 & 76 & 793\\
Present & Incorrect & 2 & 43 & 45\\
Absent & Incorrect & 138 & 239 & 377\\
Total & & 857 & 358 & 1,215\\
\midrule
\multicolumn{5}{l}{\textbf{SDBS IR--Raman ($n=282$)}}\\
Present & Correct & 200 & 17 & 217\\
Present & Incorrect & 5 & 6 & 11\\
Absent & Incorrect & 30 & 24 & 54\\
Total & & 235 & 47 & 282\\
\bottomrule
\end{tabular}
\par\smallskip
\begin{minipage}{\linewidth}\footnotesize
All test records are retained. The 17 NIST multicomponent records without fixed-workflow predictions count as incorrect. Structure identity follows the evaluation rules in Methods.
\end{minipage}
\end{table}
\FloatBarrier

\section{Peptide sequence allocation and adaptation}\label{app:peptides}

\subsection{Peptide-adapted harmonic-spectrum construction}

The fully learned peptide workflow uses peptide-adapted EquiformerV3 models for energy, direct forces, diagonal and off-diagonal Hessian blocks, dipole derivatives and polarizability derivatives. The quantum-chemical dataset contains 27,059 training, 3,023 validation and 15 test molecular identities. Energy--force training uses 432,813 optimization configurations; tensor training uses 27,059 equilibrium records. Models are initialized from the corresponding QMe14S checkpoints. The selected energy, force, diagonal-Hessian, off-diagonal-Hessian, dipole-derivative and polarizability-derivative checkpoints are epochs 11, 8, 65, 84, 87 and 86, respectively.

Harmonic inputs are generated from SMILES through conformer generation, force-field refinement, learned geometry optimization and direct tensor prediction. All 44 experimental molecules meet the predicted-force convergence threshold of 0.02~eV\,\AA$^{-1}$. Raman spectra are rendered at 785~nm with a frequency scale of 0.965 and Lorentzian FWHM of 12~\wn, on the 1,024-point 400--4,000~\wn\ grid. Experimental regions outside the reliable measurement range remain masked.

\subsection{Experimental split and HarmoNet adaptation}

The 44 experimental records comprise 19 amino-acid controls and 25 peptides. The training set contains all 19 controls and eight peptides; two peptides form the validation set, and 15 peptides form the test set. The amino-acid controls are A, R, D, N, C, Q, E, G, I, L, F, P, S, T, W, Y, V, M and K. Training peptides are GGG, GPH, KYW, LP, RKDV, SG, VYV and YFYP; validation peptides are EV and GHK. Test sequences are CF, CW, CY, EA, EY, FYY, GG, IL, LT, PHH, RKD, TR, YFY, YYF and YYY.

HarmoNet is initialized from the joint NIST/SDBS-adapted checkpoint at epoch 22 and fine-tuned for 200 epochs with batch size 8, learning rate $10^{-5}$, cosine decay to $10^{-6}$ and seed 624. Validation cosine selects epoch 199. Before/after evaluation uses the same peptide-adapted harmonic inputs, metadata, targets and masks; only the HarmoNet checkpoint changes. Main-text Fig.~\mainfigref{fig:inverse-enablement}F and Fig.~\ref{fig:peptides} use this same adapted checkpoint and test split.

\begin{figure}[ht]
\centering
\includegraphics[width=\linewidth,height=0.76\textheight,keepaspectratio]{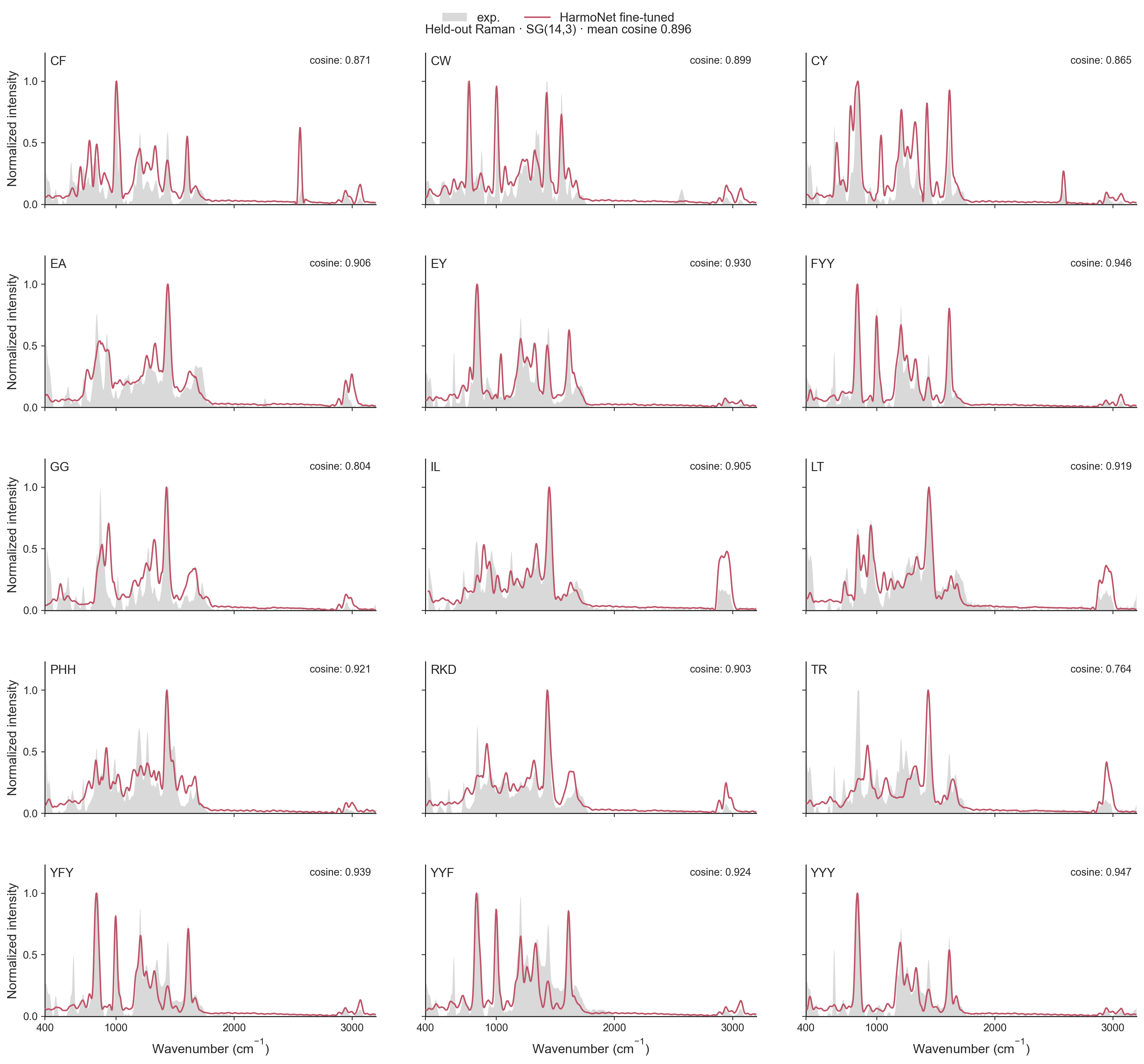}
\caption{Experimental Raman spectra and peptide-adapted HarmoNet predictions for all 15 held-out peptides using fully learned harmonic inputs. The validation-selected epoch-199 checkpoint achieves a mean cosine similarity of 0.896; the displayed range is 400--3,200~\wn. Samples are powders measured with 785-nm excitation.}\label{fig:peptides}
\end{figure}

\FloatBarrier

\clearpage
\endgroup
\end{document}